\documentclass[12pt,preprint,authoryear]{elsarticle}
\usepackage{svg}
\usepackage{amssymb}
\usepackage{amsmath}
\usepackage{amsfonts}
\usepackage{color}
\usepackage{setspace}
\usepackage{graphicx}
\usepackage{multirow}
\usepackage{epic}
\usepackage{epsfig}
\usepackage{graphics}
\usepackage{multirow}
\usepackage{subfigure} 
\usepackage{natbib} 
\usepackage{caption}
\usepackage{bm}
\usepackage{booktabs}
\usepackage[figuresright]{rotating}
\usepackage{geometry}
\usepackage{hyperref}
\usepackage{array}
\usepackage{longtable}

\usepackage{hyperref}
\hypersetup{
	colorlinks=true,    %
	linkcolor=blue,     %
	urlcolor=blue,      %
	citecolor=blue      %
}

\usepackage{bbm}
\usepackage{float}
\restylefloat{figure}
\usepackage{longtable}
\usepackage{threeparttable}
\usepackage{appendix}
\usepackage{color, xcolor}
\usepackage{tabularx}
\RequirePackage{longtable,multirow,hhline,tabularx,array}   %
\RequirePackage{flafter}    
\RequirePackage{pifont,calc}    
\RequirePackage{colortbl,booktabs}  %
\usepackage{rotating}
\usepackage{pdflscape}

\usepackage{mathrsfs}

\graphicspath{{./}{./img/}{./fig/}{./image/}{./figure/}{./picture/}}

\renewcommand{\thetable}{\arabic{table}} 
\AtBeginEnvironment{thebibliography}{%
	\setstretch{1.0} %
	\small %
}

\begin{document}
\begin{frontmatter}
	
	\title{Ultimate Forward Rate Prediction and its Application to Bond Yield Forecasting: A Machine Learning Perspective}
	
%
%
%
	
	\author[add1]{Jiawei Du}
	\ead{Jiawei.Du19@student.xjtlu.edu.cn}
	
	\author[add1]{Yi Hong$^{*}$}
	\ead{Yi.Hong@xjtlu.edu.cn}
	
	\address[add1]{School of Mathematics and Physics, Xi'an Jiaotong Liverpool University, Suzhou, Jiangsu 215123, China}
	
	\cortext[cor1]{Corresponding author. Tel: +86 (0)512 88161729. Email: Yi.Hong@xjtlu.edu.cn}
	\begin{abstract}

This study focuses on forecasting the ultimate forward rate (UFR) and developing a UFR-based bond yield prediction model using data from Chinese treasury bonds and macroeconomic variables spanning from December 2009 to December 2024. The de Kort-Vellekoop-type methodology is applied to estimate the UFR, incorporating the optimal turning parameter determination technique proposed in this study, which helps mitigate anomalous fluctuations. In addition, both linear and nonlinear machine learning techniques are employed to forecast the UFR and ultra-long-term bond yields. The results indicate that nonlinear machine learning models outperform their linear counterparts in forecasting accuracy. Incorporating macroeconomic variables, particularly price index-related variables, significantly improves the accuracy of predictions. Finally, a novel UFR-based bond yield forecasting model is developed, demonstrating superior performance across different bond maturities.

	\end{abstract}
%

\end{frontmatter}

\newpage
\section{Introduction}
The Ultimate Forward Rate (UFR) attracts significant attention from major financial institutions, such as large investment firms, pension funds, insurance companies, and long-term infrastructure investment entities. These institutions are particularly concerned with ultra-long-term interest rates. For instance, life insurance companies must evaluate the value of cash flows with maturities that extend far beyond 30 years \citep{Zhao2024}. The UFR plays a critical role in the pricing and valuation of long-term financial instruments, as well as in supporting long-term asset-liability management \citep{Christensen2021}.

Regulatory authorities also focus on the UFR, as they aim to standardize solvency assessments and mitigate pricing risks within financial markets. However, the UFR is an idealized, unobservable rate. In practice, the UFR is typically determined by the European Insurance and Occupational Pensions Authority (EIOPA). Its value is based on a combination of expected real rates and anticipated inflation rates from a group of countries, with annual updates \citep{EIOPA2022}. The extraction of the UFR from market data usually relies on specific models \citep{Du2025}. While the importance of the UFR is well recognized, prior research has primarily focused on its determination, with limited attention paid to its prediction. This study aims to bridge this gap by using machine learning techniques, incorporating a broad range of macroeconomic variables, to predict the UFR and construct bond yield term structure forecasting models based on the predicted UFR.

The UFR published annually by EIOPA is widely recognized and serves as a standard for many financial institutions. However, it has several limitations, including long intervals between updates, reliance on historical data, and an inability to capture real-time economic and policy changes. Furthermore, it may diverge from model-implied UFRs, potentially underestimating long-term interest rate risks. \cite{Christensen2021} further highlight that this divergence can lead to the overestimation of liabilities for financial institutions, such as insurance companies.

Currently, limited methods exist to estimate the UFR. Aside from the official EIOPA designation, \cite{Christensen2021} treat the level factor in the Dynamic Nelson-Siegel (DNS) model as the UFR, thus transforming the task of estimating the UFR into that of estimating the level factor within the DNS model. However, this approach also has several drawbacks, including its heavy reliance on model assumptions and issues with parameter estimation. \cite{Kort2016}  introduces a class of endogenous UFR estimation methods, known as the de Kort-Vellekoop-type methods, which are based on the Smith-Wilson method proposed by \cite{Smith2001} and the smoothness of the curve. These methods allow for the direct extraction of the UFR from market data. Despite their advantages, the "Smoothest Discount Factor" (SDF) method within the de Kort-Vellekoop-type framework has been criticized for occasionally producing negative UFR values. In response, \cite{Zhao2024} propose an improved UFR estimation method based on the SDF approach, referred to as the ZJW Improved method. This method ensures that the UFR remains positive and incorporates prior information, enhancing the original SDF method. However, the ZJW Improved method does not provide a mechanism for estimating the optimal turning parameter, as \cite{Zhao2024} only conduct a comparative analysis of different parameters.


Bond yield forecasting has garnered significant attention in academic research. Affine term structure models, such as those discussed by \cite{Ho1986}, \cite{Heath1992}, \cite{Ajello2020}, and \cite{Vayanos2021}, have been widely studied in the field of bond yield modeling. However, \cite{Koopman2010} demonstrated that these models show poor forecasting performance, often underperforming a simple random walk. Similarly, several non-affine models used by \cite{DieboldLi2006}, including Slope Regression, Fama–Bliss Forward Rate Regression, and Cochrane–Piazzesi Forward Curve Regression, fail to outperform random walk models in yield predictions. \cite{DieboldLi2006} employ the Nelson Sigel (NS) model to forecast the yield curve, and subsequent studies \cite{Yu2011}, \cite{Diebold2012}, \cite{Hevia2015} and \cite{Fernandes2019} have extended the research on DNS-type models for bond yield predictions. In particular,  \cite{Diebold2012} and \cite{Fernandes2019} integrated macroeconomic variables into these models, revealing the relationship between macroeconomic factors and bond yields across different maturities. 

Despite these advancements, research focusing specifically on forecasting ultra-long-term bond rates (UFRs) remains limited. Most existing literature concentrates on UFR determination rather than prediction. Notable contributions, such as those by \cite{Kort2016}, \cite{Christensen2021} and \cite{Zhao2024}, have explored various methodologies for deriving the UFR from market data, providing dynamic alternatives to the annual EIOPA UFR.  Among these, the de Kort-Vellekoop-type methods, which extend beyond the DNS framework and rely on the Smith-Wilson method, have proven effective in capturing the endogenous dynamics of UFRs. Once a reliable proxy for the UFR is established, several forecasting algorithms, including machine learning techniques, can be employed. Machine learning has gained widespread adoption in bond markets in recent years. For example, \cite{Bianchi2021}, \cite{Jiang2024} and \cite{Zhai2024} applied various machine learning algorithms, incorporating a wide range of macroeconomic variables, to predict bond risk premiums. Their studies highlight the nonlinear relationships between macroeconomic factors and the prediction of bond risk premiums.

This study aims to address the gap in UFR prediction and application. We combine bond yields and macroeconomic variables, employing both linear and nonlinear machine learning methods to predict UFRs. Additionally, we utilize SHAP values to analyze the interpretability of the predictive model, revealing the influence of macroeconomic factors on the forecasted outcomes. Finally, we construct a novel UFR-based bond yield forecasting model based on the predicted UFR.

Our research makes the following contributions: First, we integrate de Kort-Vellekoop-type methods to determine the optimal turning parameter within the ZJW improved method, effectively mitigating abnormal fluctuations in the UFR. Second, we apply both linear and nonlinear machine learning techniques, combined with a broad set of macroeconomic variables, to predict UFRs, addressing a gap in UFR prediction literature. Third, we uncover nonlinear relationships between input features and UFRs, particularly highlighting the superior performance of neural network models in capturing these relationships. Fourth, we demonstrate that incorporating macroeconomic variables significantly enhances forecasting accuracy, especially in nonlinear models, while their impact on linear models varies. Fifth, we perform an interpretability analysis of macroeconomic variables, clarifying their role in predicting UFRs and ultra-long-term treasury bond yields. Notably, the macroeconomic group associated with the Price Index exhibits substantial predictive power. Lastly, we propose a novel UFR-based bond yield forecasting model, which performs exceptionally well in bond yield predictions and represents the first application of UFR in forecasting the term structure of bond yields.

The remainder of this research is structured as follows: Section 2 discusses the estimation methods for UFRs and the machine learning approaches employed. Section 3 presents the data sources for Chinese Treasury bonds, the selection of macroeconomic variables, and the categorization of these variables. Section 4 reports the estimation results for UFRs, the empirical findings related to forecasting UFRs, and provides an interpretability analysis of the variables. Section 5 empirically analyzes the UFR-based bond yield forecasting model. Finally, Section 6 concludes the research.

\section{Methodology}
In this section, we will introduce methods for deriving the UFR and explore both linear and nonlinear machine learning techniques for forecasting the UFR.

\subsection{Estimation of Ultimate Forward Rates}

The  de Kort-Vellekoop-type methods proposed by \cite{Kort2016} are considered effective approaches for deriving the endogenous UFRs, building upon the Smith-Wilson method and smoothness of the curve \citep{Zhao2024}. In this section, we will discuss the Smith-Wilson method, de Kort-Vellekoop-type methods and ZJW improved method.

\subsubsection{Smith-Wilson Method}

The Smith-Wilson method is a key framework for modeling interest rate term structures, widely used in finance, especially in insurance. Developed by Bacon and Woodrow \citep{Smith2001}, it assumes a finite set of observable, liquid, risk-free fixed-income instruments with deterministic cash flows. The method requires external specification of the UFR and the convergence factor $\alpha$, with accurate estimation of the UFR being crucial due to its close relationship with $\alpha$.

Let $N$ denote the number of instruments with observed market prices $m_1, m_2, \ldots, m_N$. For an instrument with $J$ distinct maturities, the cash flows at payment dates $u_1, u_2, \ldots, u_J$ are represented as $c_{i, j}$, where $i=1, \ldots, N$ and $j=1, \ldots, J$. To ensure the linear independence of instruments, it is required that $J \geq N$. The theoretical relationship between market prices and the discount function is expressed as:

\begin{equation} \label{SW_mi}
m_i=\sum_{j=1}^J c_{i, j} \cdot P\left(u_j\right), \quad i=1, \ldots, N
\end{equation}

The Smith-Wilson methodology models the discount function $P(\tau)$ in terms of two exogenous parameters: $f_{\infty}$, the continuously compounded UFR, i.e. $f_{\infty} = \ln(1+\text{UFR})$, and $\alpha$, the convergence parameter. The discount function is expressed as:

\begin{equation} \label{SW-def}
\begin{aligned}
P(\tau)&=(1+g(\tau)) e^{-f_{\infty} \tau}\\
&=e^{-f_{\infty} \tau}\left(1+e^{f_{\infty} \tau} \sum_{i=1}^N \xi_i \sum_{j=1}^J c_{i, j} W\left(\tau, u_j\right)\right)
\end{aligned}
\end{equation}
where $g(\tau)$ is an interpolating function derived from the Wilson function. This formulation ensures smooth and coherent interpolation for observed maturities and extrapolation for longer horizons. $\xi_i(i=1, \ldots, N)$ are parameters fitted to market data, and $W\left(\tau, u_j\right)$ is the Wilson function, defined as:
\begin{equation}\label{W_and_H}
\begin{aligned}
W\left(\tau, u_j\right)&=e^{-f_{\infty}\left(\tau+u_j\right)} H\left(\tau, u_j\right)\\
H\left(\tau, u_j\right)&=\left\{\alpha \cdot \min \left(\tau, u_{j}\right)-0.5 e^{-\alpha 	\max \left(\tau, u_{j}\right)}\left(e^{\alpha \cdot \min 	\left(\tau, u_{j}\right)}-e^{-\alpha \cdot \min \left(\tau, 	u_{j}\right)}\right)\right\} .
\end{aligned}
\end{equation}
The Wilson function governs the weight matrix, which assigns weights to cash flows at different maturities, thereby enabling interpolation and extrapolation of the yield curve.

\subsubsection{De Kort-Vellekoop-Type Methods}\label{dkv}
The de Kort-Vellekoop-type methods utilizes the Smith–Wilson class of interpolating functions, framing them as the solution to a functional optimization problem. These approaches are extended to ensure that forward rates converge to a value derived from the optimization process.

We begin by presenting the SDF method.  For a given parameter $\alpha>0$, \cite{Kort2016} proposed the following optimization problem for solving endogenous UFRs:

\begin{equation}\label{ufr:opt_discount}
\arg \min _{f_{\infty}} \min _{g \in \mathcal{H}\left(f_{\infty}\right)} \int_0^{\infty}\left[g^{\prime \prime}(s)^2+\alpha^2 g^{\prime}(s)^2\right] \mathrm{d} s,
\end{equation}
where
\begin{equation}\label{ufr:opt_discount_H}
\mathcal{H}\left(f_{\infty}\right)=\left\{g \in \mathcal{C}^2\left(\mathbb{R}_{+}\right): \sum_{j=1}^T c_{i j} e^{-f_{\infty} u_j} g\left(u_j\right)= m_i-\sum_{j=1}^T c_{i j} e^{-f_{\infty} u_j}, i=1,2, \ldots, N\right\}.
\end{equation}

\hyperref[opt_discount]{Eq. (\ref{ufr:opt_discount})} requires us to determine an optimal value for $f_{\infty}$ that minimizes the objective function. 
Given the observed market data, we define an $N \times J$ cash flow matrix $\mathbf{C}$, a $J \times J$ matrix $\mathbf{W}$ as:

\begin{equation}
\begin{aligned}
\mathbf{C}_{i j}&=c_{i j}, \quad i=1,2, \ldots, N ; \quad j=1,2, \ldots, J\\
\mathbf{W}_{i j}&=W_\alpha\left(u_i, u_j\right), \quad i, j=1,2, \ldots, J.
\end{aligned}
\end{equation}
Furthermore, let
\begin{equation}
\mathbf{m}=( \begin{array}{llll}
m_1 & m_2 & \cdots & m_N
\end{array} )^{\top}, \mathbf{D}_{i j}^f=e^{-f_{\infty} u_j} \mathbf{1}_{i=j}, \quad \mathbf{U}_{i j}=u_j \mathbf{1}_{i=j}.
\end{equation}
where $\mathbf{m}$ represents observed market bond prices vecter. \cite{Kort2016} concluded that the optimal $f_{\infty}$ for \hyperref[opt_discount]{Eq. (\ref{ufr:opt_discount})} should satisfy the following first-order condition:
\begin{equation}
\begin{aligned}
& \left(\mathbf{m}-\mathbf{C D}^f_{\infty} \mathbf{e}\right)^T\left(\mathbf{C D}^f_{\infty} \mathbf{W D}^f_{\infty} \mathbf{C}^T\right)^{-1} \mathbf{C D}^f_{\infty} \mathbf{U} \\
& \quad \times\left(\mathbf{e}+\mathbf{W D}^f_{\infty} \mathbf{C}^T\left(\mathbf{C D}^f_{\infty} \mathbf{W D}^f_{\infty} \mathbf{C}^T\right)^{-1}\left(\mathbf{m}-\mathbf{C D}^f_{\infty} \mathbf{e}\right)\right)=0,
\end{aligned}
\end{equation}
where $\mathbf{e}=( \begin{array}{llll}
1 & 1 & \cdots & 1
\end{array} )^{\top}$. If the cashflow matrix $\mathbf{C}$ is invertible this simplifies to
\begin{equation} \label{foc_dkv}
\sum_{i=1}^T \sum_{j=1}^T\left(u_i \pi_i e^{f_{\infty} u_i}\right)\left[\mathbf{W}^{-1}\right]_{i j}\left(\pi_j e^{f_{\infty} u_j}-1\right)=0
\end{equation}
with $\pi_i=\sum_{j=1}^T\left[\mathbf{C}^{-1}\right]_{i j} m_j$.

By adopting this approach, we not only obtain a smoother curve in comparison to that generated by the original Smith-Wilson method, but also derive an endogenous UFR. The core of this method lies in the optimization of the discount factor's smoothness, which is henceforth referred to as SDF method.

Two alternative methods for determining the UFR are the Smoothest Forward Rate (SFR) method and the Smoothest Yield Curve (SYC) method. These methods obtain the converging UFR by solving \hyperref[opt_frd]{Eq. (\ref{opt_frd})} and \hyperref[opt_yield]{Eq. (\ref{opt_yield})}, respectively.

\begin{equation}\label{opt_frd}
\min _{g \in \mathcal{H}^f } \int_0^{\infty}\left[g^{\prime \prime}(s)^2+\alpha^2 g^{\prime}(s)^2\right] \mathrm{d} s,
\end{equation}
\begin{equation}\label{opt_yield}
\min _{g \in \mathcal{H}^y } \int_0^{\infty}\left[g^{\prime \prime}(s)^2+\alpha^2 g^{\prime}(s)^2\right] \mathrm{d} s,
\end{equation}
where
\begin{equation}\label{opt_frd_H}
\mathcal{H}^f =\left\{g \in \mathcal{C}^2\left(\mathbb{R}_{+}\right): \sum_{j=1}^T c_{i j} e^{-\int_0^{u_j} g(s) d s}=m_i, i=1,2, \ldots, N\right\},
\end{equation}

\begin{equation}\label{opt_yield_H}
\mathcal{H}^y =\left\{g \in \mathcal{C}^2\left(\mathbb{R}_{+}\right): \sum_{j=1}^T c_{i j} e^{-u_j g\left(u_j\right)}=m_i, i=1,2, \ldots, N\right\}.
\end{equation}

The two estimated asymptotic UFR can be expressed as linear combinations of yields at different maturities. By incorporating the Wilson function into the solving process, the UFR can be derived. The detailed solution process is provided in Appendix \ref{ufr_sfr_syc}, and further details can be found in \cite{Kort2016}.

\subsubsection{ZJW Improved Method: Non-Negativity Constrained de Kort-Vellekoop Method with Prior Information}
\cite{Zhao2024} further developed an enhanced version of the SDF method with constraints. This method incorporates prior knowledge about the UFR into the framework and, on the other hand, includes the dynamic determination of $\alpha$, ensuring that the UFR remains positive. The approach introduces a new optimization problem:
\begin{equation}
\arg \min _{f_{\infty}} \min _{g \in \mathcal{H}\left(f_{\infty}\right)}\left[\int_0^{\infty}\left[g^{\prime \prime}(s)^2+\alpha^2 g^{\prime}(s)^2\right] \mathrm{d} s+\frac{\lambda}{2} \alpha^3\left(f_{\infty}-f_{\text {prior }}\right)^2\right]
\end{equation}
where $\mathcal{H}\left(f_{\infty}\right)$ is as defined in \hyperref[opt_discount_H]{Eq. (\ref{ufr:opt_discount_H})} , $f_{\text {prior }}>0$ represents the prior knowledge about the UFR (the $f_{\text {prior }}$ was set to 4.5\% \citep{Zhao2024}), and $\lambda>0$ is a tuning parameter.
Thus, the first-order condition in equation \hyperref[foc_dkv]{Eq. (\ref{foc_dkv})} becomes:
\begin{equation}\label{foc_zjw}
\sum_{i=1}^T \sum_{j=1}^T\left(u_i \pi_i e^{f_{\infty} u_i}\right)\left[\mathbf{W}_\alpha^{-1}\right]_{i j}\left(\pi_j e^{f_{\infty} u_j}-1\right)+\lambda\left(f_{\infty}-f_{\text {prior }}\right)=0 .
\end{equation}

Before solving \hyperref[foc_zjw]{Eq. (\ref{foc_zjw})}, the ZJW improved method requires determining an optimal value for $\alpha$, as per the EIOPA approach. This process involves two steps: the first step is to derive the feasible region for $\alpha$, and the second step is to select the optimal value of $\alpha$. The definition of the feasible region for $\alpha$ in the first step is as follows:
\begin{equation}
	\begin{aligned}
		 \tilde{\mathcal{A}}:&=\left\{\alpha: \alpha>0, \sum_{i=1}^T \sum_{j=1}^T\left(u_i \pi_i\right)\left[\mathbf{W}_\alpha^{-1}\right]_{i j}\left(\pi_j-1\right)-\lambda f_{\text {prior }}<0\right\} \\
		 \tilde{\mathcal{B}}:&=\left\{\alpha: \alpha \in \tilde{\mathcal{A}}, \alpha \geq \alpha_{\min },\left|\tilde{f}^\alpha(\mathrm{CP})-\tilde{f}_{\infty}^\alpha\right| \leq \tau\right\}
	\end{aligned}
\end{equation}
where $f_{\infty}^\alpha$ denotes the solution to the first-order condition in \hyperref[foc_zjw]{Eq. (\ref{foc_zjw})} given $\alpha$, and $f^\alpha(\cdot)$ represents the forward curve generated by the Smith-Wilson method given $\alpha$ and $f_{\infty}=f_{\infty}^\alpha$.

Subsequently, we proceed to select an optimal value for $\alpha$, which corresponds to Step 2:
\begin{equation}
\alpha^*=\inf \mathcal{B} .
\end{equation}

Finally, we can solve the first-order condition, as represented by  \hyperref[foc_zjw]{Eq. (\ref{foc_zjw})}, to obtain $f_{\infty}$.

One advantage of the ZJW improved method is its ability to mitigate the extreme values of $f_{\infty}$ that arise from the SDF method. However, it does not explicitly specify the optimal value for the turning parameter $\lambda$. In this paper, we propose that the optimal $\lambda$ be determined by minimizing the sum of squared errors between $\mathrm{UFR_{ZJW}}$ and both $\mathrm{UFR_{SFR}}$ and $\mathrm{UFR_{SYC}}$. The $\mathrm{UFR_{SFR}}$ and $\mathrm{UFR_{SYC}}$ series, obtained using the SFR and SYC methods outlined in Section \ref{dkv}, serve as the benchmarks for comparison. Let $\mathrm{UFR_{SFR}}$ , $\mathrm{UFR_{SYC}}$ and $\mathrm{UFR_{ZJW}}$ represent the three time series, where $\mathrm{UFR_{ZJW}}$ is generated using ZJW improved method with the parameter $\lambda$. We aim to minimize the following objective function by adjusting $\lambda$ :
\begin{equation}
\lambda^*=\arg \min _\lambda\left[\sum_{t=1}^T\left(\mathrm{UFR_{ZJW}}(t, \lambda)-\mathrm{UFR_{SFR}}(t)\right)^2+\sum_{t=1}^T\left(\mathrm{UFR_{ZJW}}(t, \lambda)-\mathrm{UFR_{SYC}}(t)\right)^2\right].
\end{equation}
where $\mathrm{UFR_{ZJW}}(t, \lambda)$ denotes the value of the time series $\mathrm{UFR_{ZJW}}$ at time $t$, computed using ZJW improved method and parameter $\lambda$.  $\mathrm{UFR_{SFR}}(t)$ and $\mathrm{UFR_{SYC}}(t)$ are the values of the time series $\mathrm{UFR_{SFR}}$ and $\mathrm{UFR_{SYC}}$ at time $t$, respectively. $T$ is the length of the time series. Once $\lambda^*$ is determined, we obtain $\mathrm{UFR_{ZJW}}\left(t, \lambda^*\right)$.

\subsection{Forecasting Methods}
In this section, we present a comprehensive overview of various forecasting methods. These methods encompass econometric techniques such as Ordinary Least Squares (OLS), as well as linear machine learning approaches including Principal Component Regression (PCR), Partial Least Squares (PLS), Ridge, Lasso, and Elastic Net (EN). Additionally, we examine nonlinear machine learning methods, including Regression Trees (RT), Gradient-Boosting Regression Trees (GBRT), Extreme Gradient Boosting (XGBoost), and Neural Networks (NN).

\subsubsection{Ordinary Least Squares}
In the OLS model, the dependent variable is the $\operatorname{\Delta UFR}$ (first-order differences, hereafter), while the independent variables consist of China treasury bond yields (first-order differences, hereafter), with maturities ranging from 1 to 50 years. The relationship between the dependent and independent variables is defined as follows:
\begin{equation} \label{ols1}
\operatorname{\Delta UFR}_{t+1} =c + a_{1} \Delta y_{t}^{(1)} + a_{2} \Delta y_{t}^{(2)} +\cdots+  a_{m} \Delta y_{t}^{(n)} + \varepsilon_{t},
\end{equation}
where $c$ is a constant, the error term, $\varepsilon_{t}$, is assumed to satisfy the standard regression assumptions, including having a zero mean and being uncorrelated with the independent variables. $\operatorname{\Delta UFR}_{t+1} = \operatorname{UFR}_{t+1}-\operatorname{UFR}_{t}$, and $\Delta y_t^{(n)} = y_{t}^{(n)} - y_{t-1}^{(n)}$ represents the first-order difference of the yield of the treasury bond with an $n$-year maturity\footnote{The first differences of the UFR and bond yield series are taken since \hyperref[table:unitroot]{Table \ref{ufr:table:unitroot}} shows that these series are non-stationary and exhibit a unit root, which can impact the predictive performance of OLS and other linear machine learning models.}. 

In the second model, , we incorporate a set of macroeconomic factors into the forecast. The objective is to investigate the predictive power of the regression model when these macroeconomic variables are included. The model, which includes macroeconomic variables of different categories, is defined as follows:
\begin{equation} \label{ols2}
\operatorname{\Delta UFR}_{t+1} = c + a_{1} \Delta y_{t}^{(1)} + a_{2} \Delta y_{t}^{(2)} +\cdots+  a_{m} \Delta y_{t}^{(n)} + b_{1} f_{t}^{(1)} + b_{2} f_{t}^{(2)} +\cdots+  b_{k} f_{t}^{(k)} + \varepsilon_{t},
\end{equation}
where $f_t^{(k)}$ denotes the $k$-th macroeconomic variable. We refer to \cite{Jiang2024} for the OLS model, where only a representative subset of macroeconomic variables is selected. Specifically, we extract the first principal component from each of the 13 major categories of macroeconomic variables as features. This choice is driven by the inherent limitations of OLS in handling a large number of independent variables. For the other methods, all macroeconomic variables are selected, with $k$ set to 105. Detailed classifications and selections of macro variables are provided in Section \ref{ufr:macro_selection} and Appendix \ref{ufr:abbr_macro}.

\subsubsection{Principal Component Regression and Partial Least Squares}
To address potential multicollinearity among the predictors, particularly the yield rates of government bonds with varying maturities, and the high dimensionality of our feature set, we apply PCR and PLS. The technique of PCR effectively mitigates multicollinearity by transforming the predictors into orthogonal components, which simplifies the regression model's structure. We define $u_t$ and $v_t$ as the principal components (PCs) of forward rates and macroeconomic variables, respectively. The model specifications presented in \hyperref[ols1]{Eq. (\ref{ols1})} and \hyperref[ols2]{Eq. (\ref{ols2})} are revised as follows:
\begin{equation} 
\begin{gathered} \operatorname{\Delta UFR}_{t+1}^{(n)} = \Lambda_1 + A u_t + \eta_t, \quad \text{with} , u_t = \Gamma_1 \Delta y_t + p, \\
\ \operatorname{\Delta UFR}_{t+1}^{(n)} = \Lambda_2 + A u_t + C v_t + e_t, \quad \text{with} , v_t = \Gamma_2 f_t + q, \end{gathered} \end{equation}
where $A$ and $C$ represent the coefficient matrices for the PCs derived from forward rates and macroeconomic factors, respectively, while $\Gamma_1$ and $\Gamma_2$ are the corresponding transformation matrices. In contrast, PLS extracts components by leveraging the joint distribution of the predictors, which enables it to capture the underlying data structure more effectively, especially in the presence of highly correlated variables. This characteristic is particularly important for forecasting the UFR.

\subsubsection{Penalized Regressions: Ridge, Lasso, and Elastic Net}
Given the potential multicollinearity among the features used for forecasting, particularly between treasury bond yields at different maturities, we employ penalized regression techniques, such as Ridge, Lasso, and Elastic Net, to address this issue and reduce overfitting. These methods are commonly applied in financial data analysis to improve predictive accuracy in the presence of noise. By incorporating a penalty term into the OLS loss function, they minimize the influence of weak or irrelevant predictors, thereby enhancing the model's robustness and performance.

Ridge regression applies an $L_2$ penalty, shrinking coefficients toward zero without eliminating them. Lasso, on the other hand, uses an $L_1$ penalty that can reduce some coefficients to exactly zero, effectively performing variable selection. Elastic Net combines both penalties, providing a balance between Ridge and Lasso's advantages. The general form of penalized regressions is expressed as:
\begin{equation}
\mathcal{L}(\Theta ; \cdot)=\underbrace{\mathcal{L}_{O L S}(\theta)}_{\text {Loss Function }}+\underbrace{\phi(\beta ; \cdot)}_{\text {Penalty Term }}
\end{equation}
where $\mathcal{L}_{O L S}(\theta)$ is the OLS loss function, and $\phi(\beta)$ is the penalty term, which varies for each method:
\begin{equation}
\phi(\beta)= \begin{cases}\lambda \sum_{j=1}^p \beta_j^2 & \text { Ridge } \\ \lambda \sum_{j=1}^p\left|\beta_j\right| & \text { Lasso } \\ \lambda \mu \sum_{j=1}^p\left|\beta_j\right|+\frac{\lambda(1-\mu)}{2} \sum_{j=1}^p \beta_j^2 & \text { Elastic Net }\end{cases},
\end{equation}
where $\lambda$ and $\mu$ are hyperparameters controlling the shrinkage and regularization level. These techniques help improve model robustness by selecting the most relevant variables, thereby reducing overfitting and enhancing generalization.

\subsubsection{Ensemble Regression Trees: Regression trees, Gradient-Boosting Regression Trees, Extreme Gradient Boosting}
Ensemble methods, particularly RT, GBRT, and XGBoost, are powerful techniques for modeling complex nonlinear relationships in forecasting tasks.

A RT partitions the feature space into regions, with each region corresponding to a predicted value, typically the mean of the target variable. The tree is recursively split by selecting thresholds that minimize the variance within each region. Mathematically, the model can be expressed as:
\begin{equation}
\hat{\mathrm{UFR}}=\sum_{k=1}^K c_k \mathbb{I}\left(x \in R_k\right),
\end{equation}
where $c_k$ is the mean target value in leaf $R_k$, and $\mathbb{I}\left(x \in R_k\right)$ is an indicator function for whether $x$ falls into region $R_k$.

GBRT improves upon individual regression trees by sequentially fitting new trees to the residual errors of the previous model. Each tree is trained to minimize the residual sum of squares using gradient descent on a specified loss function. The update rule for the $m$-th iteration is:
\begin{equation}
F_m(x)=F_{m-1}(x)+\eta h_m(x),
\end{equation}
where $h_m(x)$ is the new tree, $F_{m-1}(x)$ is the current model, and $\eta$ is the learning rate.

XGBoost enhances GBRT with additional regularization to control model complexity. The objective function includes both the loss term and a regularization term that penalizes large trees:
\begin{equation}
\mathcal{L}=\sum_{i=1}^n \mathcal{L}\left({\mathrm{UFR_{i}}}, \hat{\mathrm{UFR_{i}}}\right) +\sum_{k=1}^K \Omega\left(f_k\right),
\end{equation}
where $\Omega\left(f_k\right)=\gamma T+\frac{1}{2} \lambda \sum_{j=1}^T w_j^2$ is the regularization term, $T$ is the number of leaves, and $w_j$ is the weight of leaf $j$. This formulation allows XGBoost to better handle overfitting and efficiently learn from large datasets.

\subsubsection{Neural Networks}
NNs are widely utilized nonlinear machine learning methods within the broader domain of supervised learning techniques. This study follows the methodology presented in \cite{Bianchi2021} and employs feed-forward networks, specifically multilayer perceptrons (MLP). We investigate four distinct types of neural networks: Yields-Only-Net, Hybrid-Net, Double-Net, and Group ensemble net. These models are examined with different configurations, including varying numbers of hidden layers and nodes in each layer\footnote{L2 regularization was incorporated into the neural networks to mitigate overfitting and improve the model's generalization capability.}.

\begin{figure}[H]
	\centering
	\includegraphics[width=0.7\textwidth]{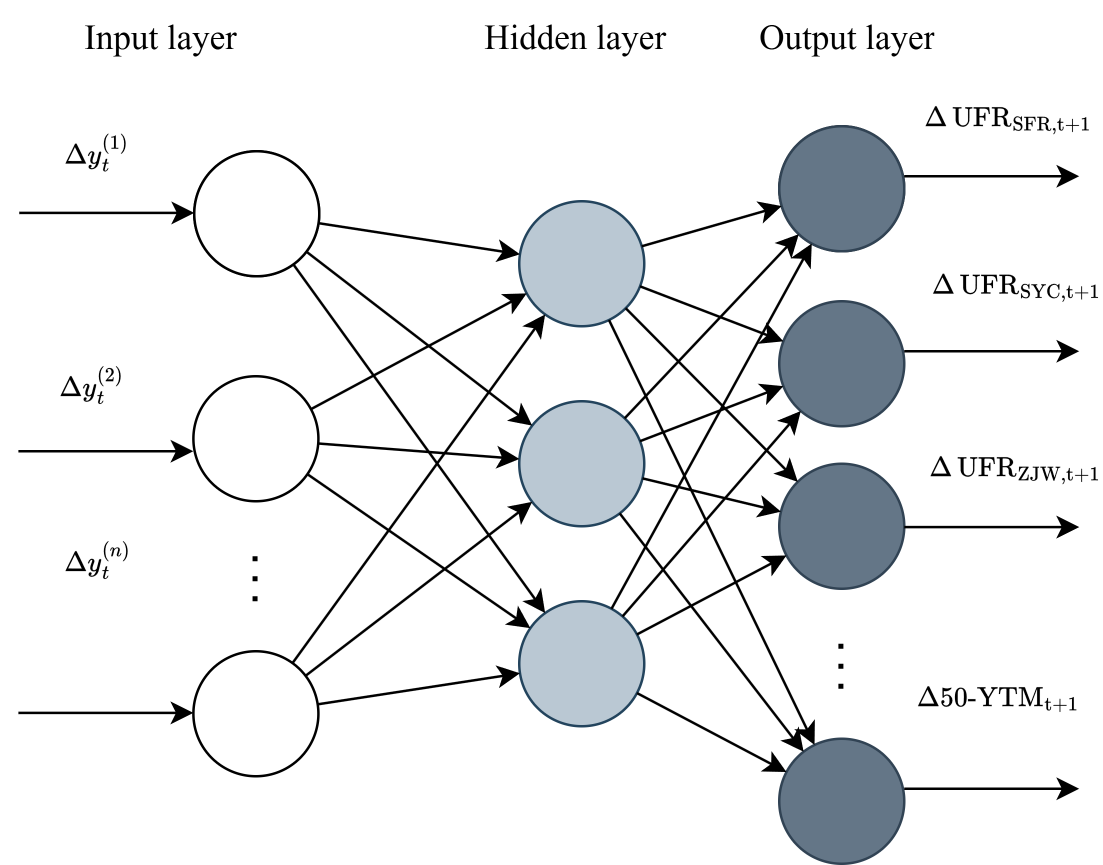}
	\caption{Yield-Only-Net.\\ \footnotesize{Note: The input features of Yield-Only-Net consist solely of the first-order differences of bond yields, denoted as $\Delta y_{t}$.}}  
	\label{NNs_yield}
\end{figure}

\begin{figure}[H]
	\centering
	\subfigure[Yield-Macro net.]{
	\includegraphics[width=0.48\textwidth]{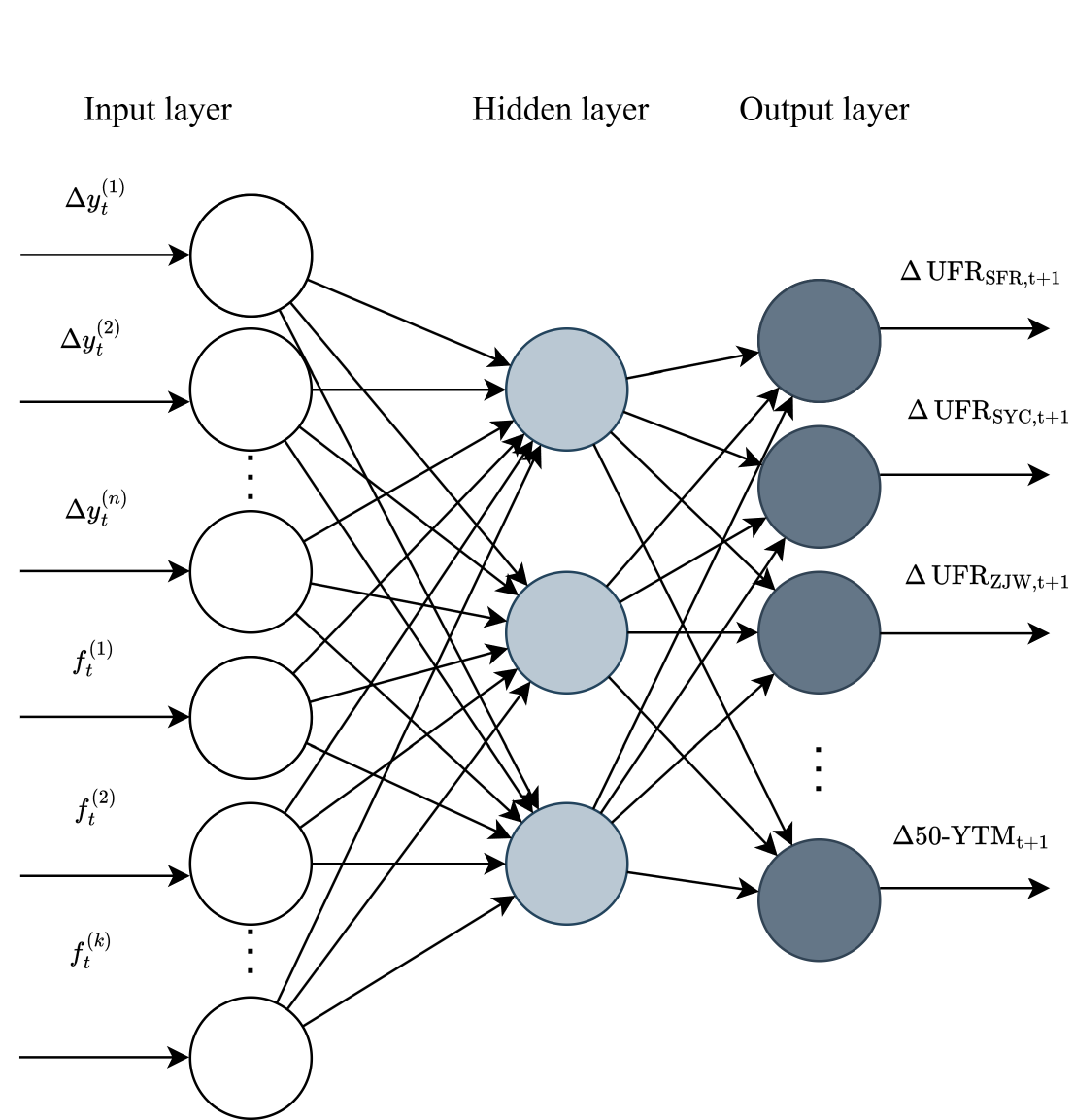}
	}
	\hspace{-3mm}
	\subfigure[Hybrid net.]{
		\includegraphics[width=0.48\textwidth]{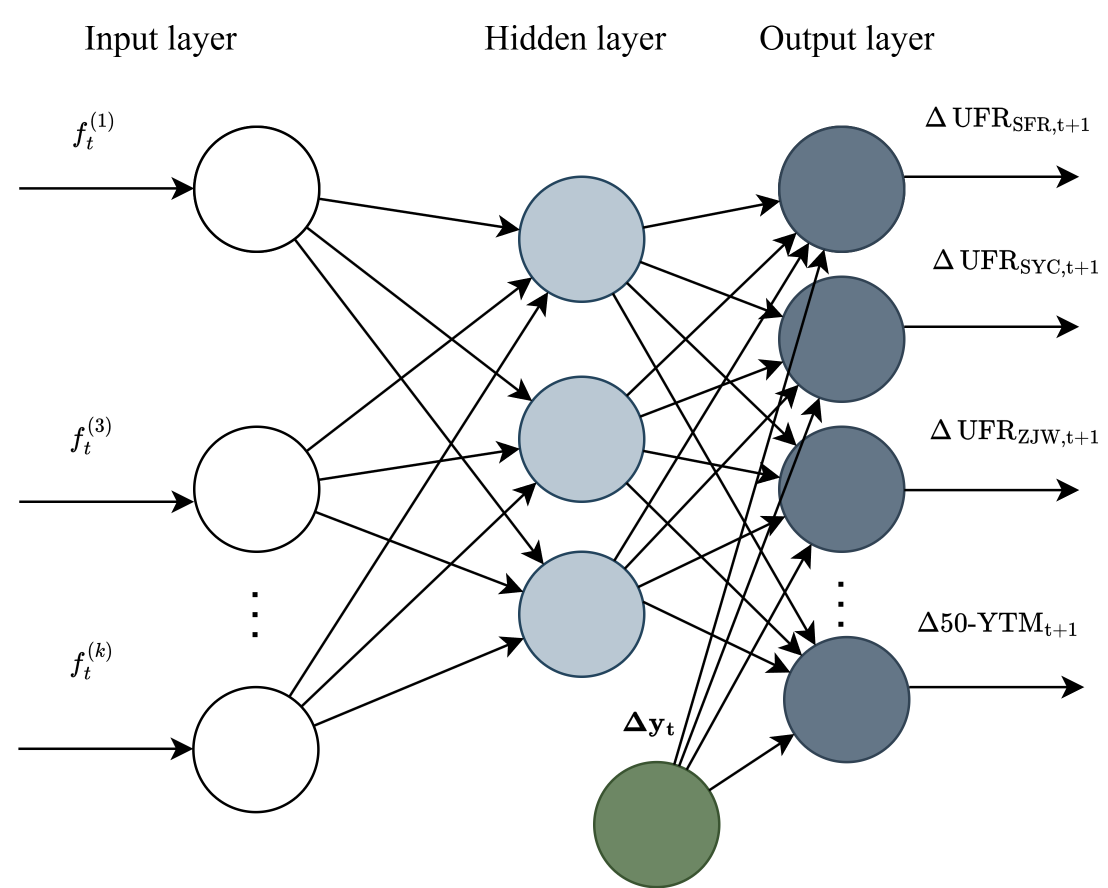}
	}
	\vspace{3mm}
	\subfigure[Double net.]{
		\includegraphics[width=0.48\textwidth]{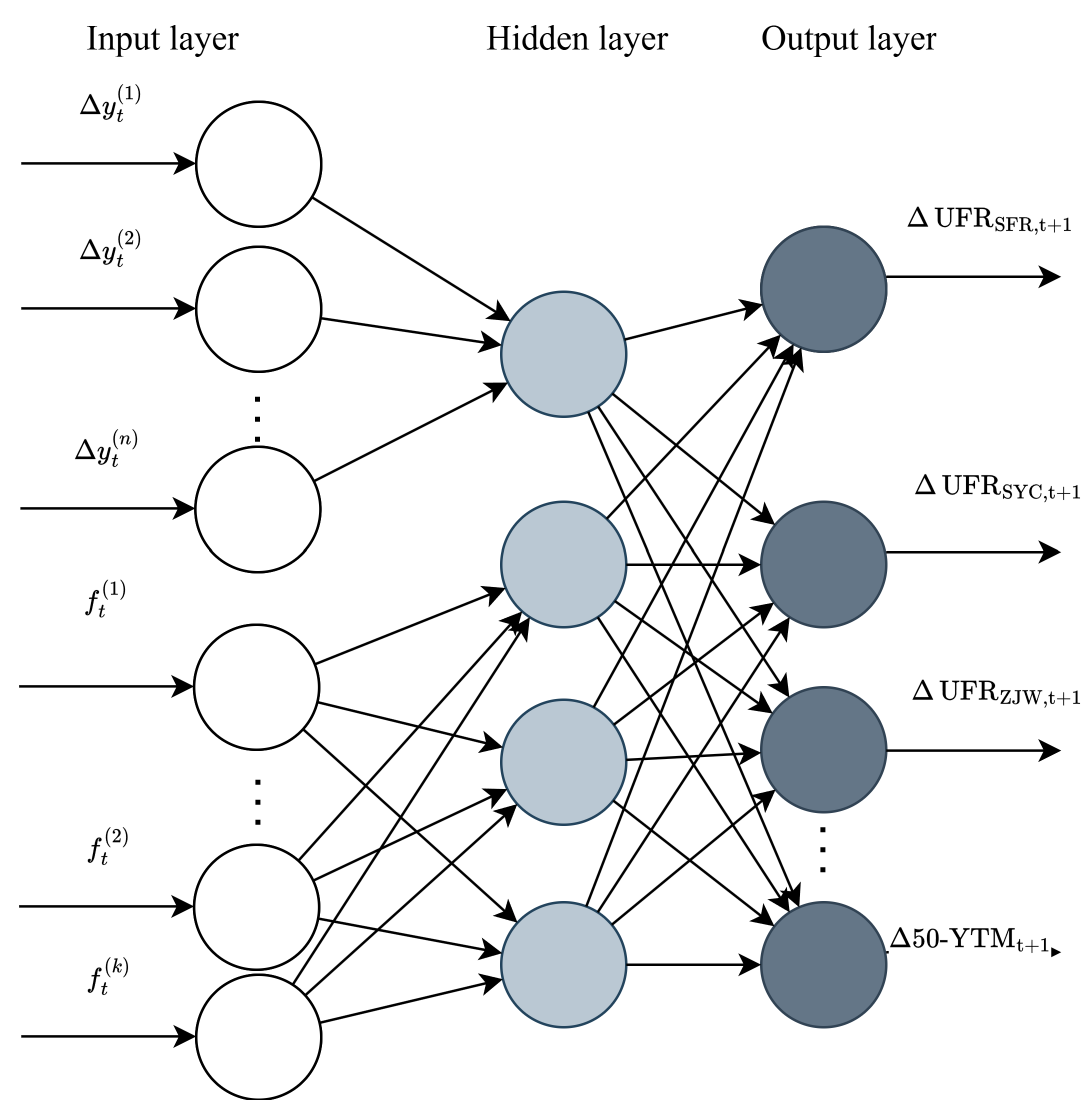}
	}
	\hspace{-3mm}
	\subfigure[Group ensemble net.]{
		\includegraphics[width=0.48\textwidth]{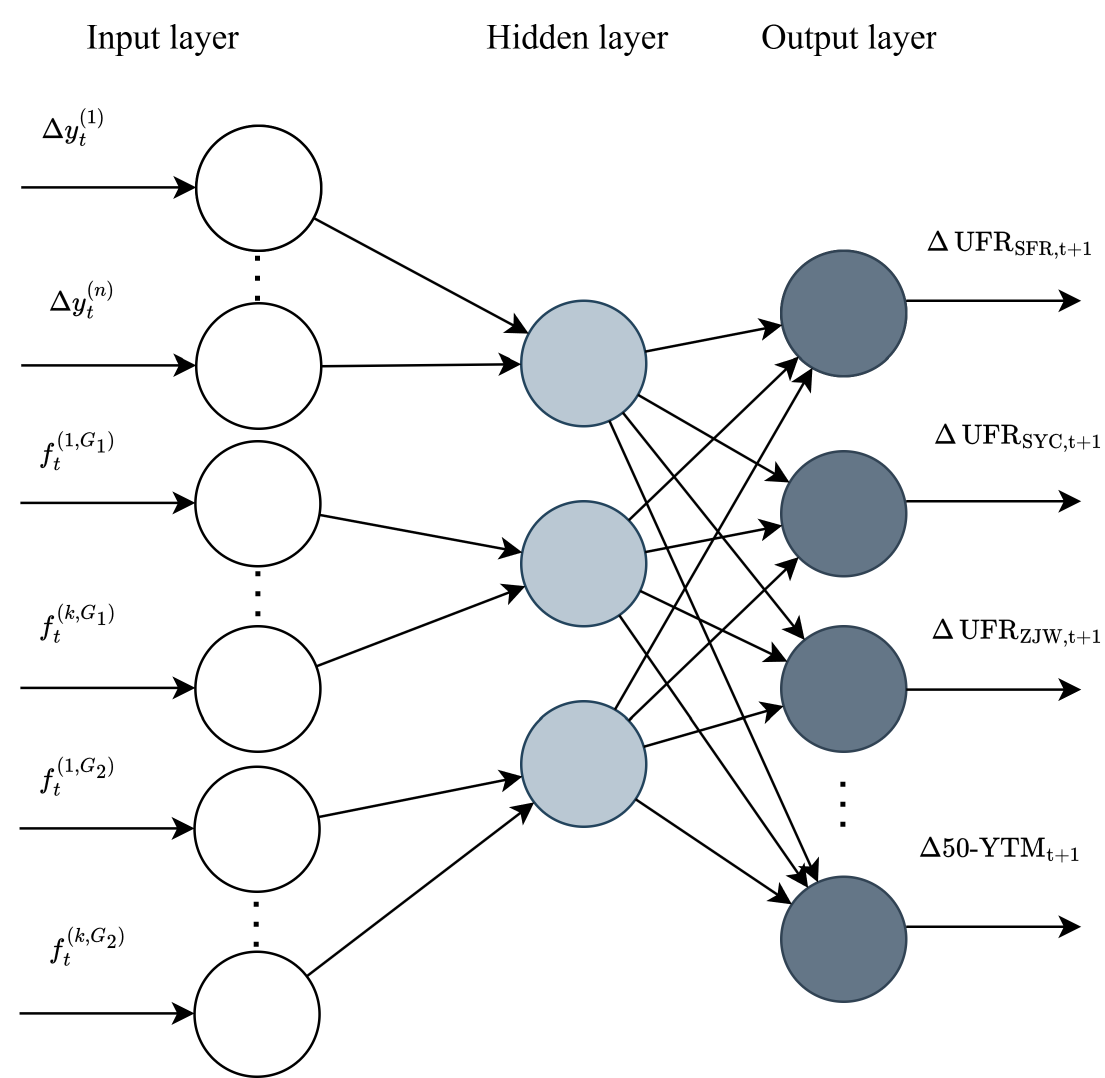}
	}
	\caption{Neural networks with yields and macros.\\ \footnotesize{The input features of Yield-Macro Net, Hybrid Net, Double Net, and Group Ensemble Net include both the first-order differences of bond yields, $\Delta y_{t}$, and macroeconomic variables. Among these, only the Hybrid Net directly combines $\Delta y_{t}$ linearly with the output layer, while the remaining networks model both $\Delta y_{t}$ and macroeconomic variables through nonlinear relationships.}}
	\label{NNs_yield_macro}
\end{figure}

The Yields-only-Net utilizes only bond yields of various maturities (first-order differences, hereafter) as input, while the Yield-Macro-Net incorporates a broader set of macroeconomic variables, as shown in \hyperref[NNs_yield]{Fig. \ref{NNs_yield}} and \hyperref[NNs_yield_macro]{Fig. \ref{NNs_yield_macro}(a)}. The remaining three networks also integrate macroeconomic variables, exhibiting more varied structures. In the Hybrid-Net, a broader set of macroeconomic variables is combined nonlinearly through hidden layers, while bond yields are combined linearly in the output layer, as illustrated in \hyperref[NNs_yield_macro]{Fig. \ref{NNs_yield_macro}(b)}.

The Double-Net, depicted in \hyperref[NNs_yield_macro]{Fig. \ref{NNs_yield_macro}(c)}, integrates two distinct networks at the output layer: one network trains on bond yields, and the other trains on macroeconomic variables. This architecture allows the network to separately capture the nonlinear relationships between macroeconomic factors and bond yields.

The final network, Group Ensemble Net (GN-Net), follows the model framework proposed by \cite{Jiang2024} and \cite{Zhai2024}. This network divides the input features into different groups, such as the Price index Group and the Interest rates Group, along with 13 other macroeconomic variable groups, and a separate bond yield group. Each group corresponds to a distinct network, and the outputs are ensembled at the output layer level, as shown in \hyperref[NNs_yield_macro]{Fig. \ref{NNs_yield_macro}(d)}.

\section{Data}

\subsection{China Treasury Bond Yields} \label{china_b_yields}
This study utilizes monthly data on Zero-Coupon China Treasury bonds provided by the China Central Depository \& Clearing Co., Ltd. The dataset covers bond maturities of 1, 2, 3, 5, 7, 10, 15, 20, 30, 40, and 50 years, spanning from December 31, 2009, to December 31, 2024. The analysis begins on December 4, 2009, as yield data for the 40-year and 50-year bonds were unavailable before this date.
\begin{figure}[H]
	\centering
	\includegraphics[width=0.8\textwidth]{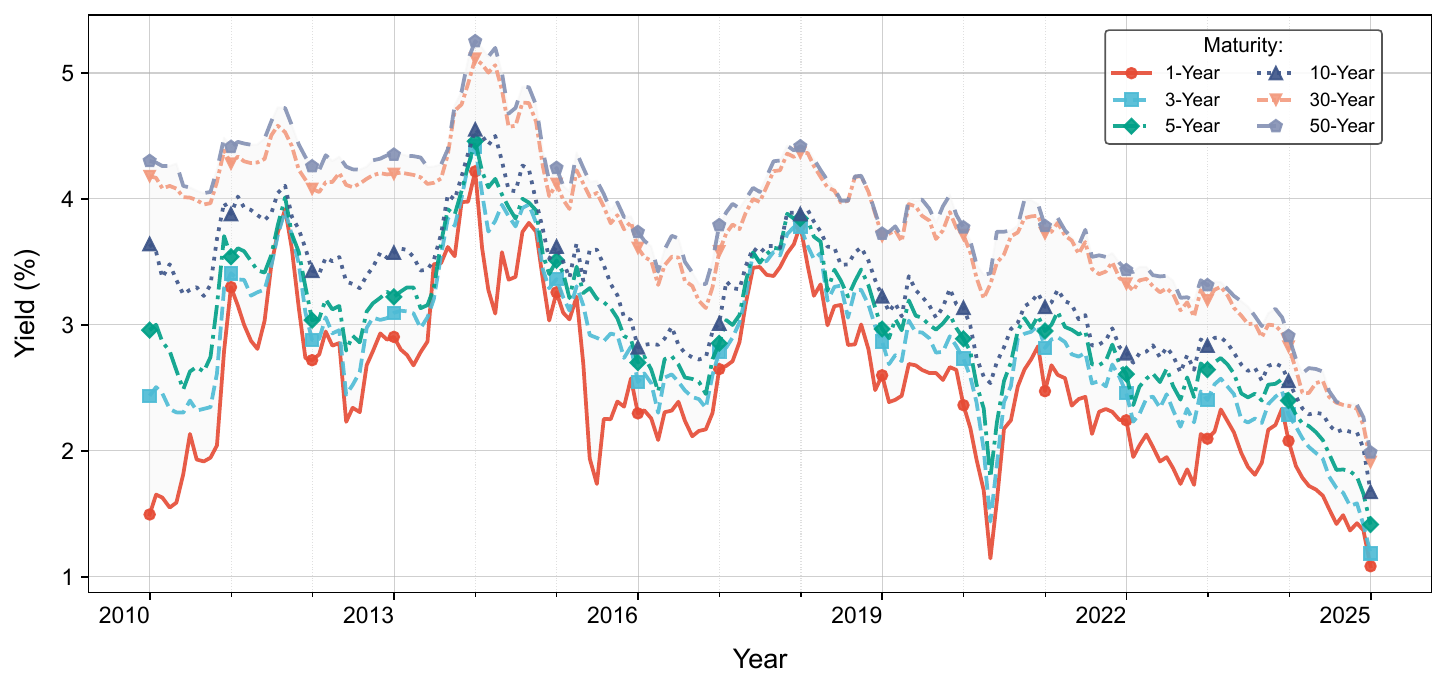}
	\caption{Dynamics of Term Structure of Treasury Bond Yields in China.\\ \footnotesize{Note: This figure illustrates the yield curves with selected maturities of 1, 3, 5, 10, 30, and 50 years from December 31, 2009, to December 31, 2024.}}  
	\label{ufr:bond_surface}
\end{figure}

To construct the yield curve shown in Fig. \ref{ufr:bond_surface}, bonds representing short-term (1-year), medium-term (3-year and 5-year), and long-term (10-year, 30-year, and 50-year) maturities were selected. The figure illustrates several key trends. From early 2011 to late 2013, yields remained high across all maturities, reflecting rising inflation and expectations of tightening monetary policies to control inflation. In 2013, the "money shortage" crisis, driven by stable economic growth, increased capital demand, and shifts in monetary policy and regulation, led to significant liquidity constraints \citep{Bai2022, Du2025}.

From early 2015 to late 2016, bond yields experienced substantial volatility, coinciding with the Chinese stock market crash \cite{Umar2021}. Between early 2020 and late 2021, yields sharply declined due to the accommodative monetary policies aimed at mitigating the economic impact of the COVID-19 pandemic. Since 2023, yields across all maturities have followed a consistent downward trend, during which China's real estate sector fell into a slump, and deflationary expectations regarding the economy intensified.

An additional key observation is that as bond maturities lengthen, the yield curve shifts upward with reduced volatility. This suggests that UFR, reflecting ultra-long-term interest levels, will exhibit lower volatility while remaining elevated.

\begin{table}[H]
	\caption{\\Correlation Matrix of Treasury Bond Yields with Different Maturities}
	\setlength{\tabcolsep}{8pt}
	\renewcommand{\arraystretch}{0.8}
	\resizebox{\textwidth}{!}{
		\begin{tabular}{@{}clllllllllll@{}}
			\toprule
			\multicolumn{1}{l}{\textbf{Maturity (years)}} &
			\multicolumn{1}{c}{\textbf{1}} &
			\multicolumn{1}{c}{\textbf{2}} &
			\multicolumn{1}{c}{\textbf{3}} &
			\multicolumn{1}{c}{\textbf{5}} &
			\multicolumn{1}{c}{\textbf{7}} &
			\multicolumn{1}{c}{\textbf{10}} &
			\multicolumn{1}{c}{\textbf{15}} &
			\multicolumn{1}{c}{\textbf{20}} &
			\multicolumn{1}{c}{\textbf{30}} &
			\multicolumn{1}{c}{\textbf{40}} &
			\multicolumn{1}{c}{\textbf{50}} \\ \midrule
			\textbf{1}  &        &        &        &        &        &        &        &        &        &        &  \\
			\textbf{2}  & 0.99** &        &        &        &        &        &        &        &        &        &  \\
			\textbf{3}  & 0.96** & 0.99** &        &        &        &        &        &        &        &        &  \\
			\textbf{5}  & 0.92** & 0.96** & 0.99** &        &        &        &        &        &        &        &  \\
			\textbf{7}  & 0.89** & 0.94** & 0.97** & 0.99** &        &        &        &        &        &        &  \\
			\textbf{10} & 0.84** & 0.89** & 0.93** & 0.97** & 0.98** &        &        &        &        &        &  \\
			\textbf{15} & 0.82** & 0.86** & 0.90** & 0.94** & 0.97** & 0.98** &        &        &        &        &  \\
			\textbf{20} & 0.80** & 0.85** & 0.89** & 0.93** & 0.95** & 0.98** & 0.99** &        &        &        &  \\
			\textbf{30} & 0.80** & 0.84** & 0.88** & 0.92** & 0.95** & 0.97** & 0.99** & 0.99** &        &        &  \\
			\textbf{40} & 0.79** & 0.84** & 0.88** & 0.92** & 0.95** & 0.97** & 0.99** & 0.99** & 1.00** &        &  \\
			\textbf{50} & 0.79** & 0.84** & 0.87** & 0.91** & 0.95** & 0.97** & 0.98** & 0.99** & 1.00** & 1.00** &  \\ \bottomrule
	\end{tabular}}
	\begin{tablenotes}
		\fontsize{10}{10}\selectfont
		\item[1] Note: A double asterisk (**) following the correlation coefficient indicates statistical significance at the 0.05 level ($p < 0.05$) based on the significance test of the Pearson correlation coefficient. 
	\end{tablenotes}
	\label{ufr:table:corr_yield}
\end{table}

The correlation matrix of bond yields for different maturities is presented in \hyperref[table:corr_yield]{Table \ref{ufr:table:corr_yield}}. \hyperref[table:corr_yield]{Table \ref{ufr:table:corr_yield}} further demonstrates that yields for bonds with different maturities are highly correlated, with stronger correlations observed for bonds with shorter maturities. Consequently, the UFR is expected to exhibit similar characteristics, showing a high correlation with the long-term bond yield curve.

\subsection{Macroeconomic Factors} \label{ufr:macro_selection}
The UFR represents the expected long-term limit of interest rates in a stable state and is closely tied to the yield on long-term treasury bonds. A well-established relationship exists between macroeconomic variables and treasury bond yields, which has been extensively studied. Previous research has primarily focused on the term structure of yields, bond risk premiums, bond volatility, and the impact of policy shocks on bond yields. These studies have explored the relationship between bond yields and inflation \citep{Fan2010, Fan2012, Shang2023}, investment \citep{Diebold2012, Chionis2014, Fernandes2019}, international trade \citep{Yan2015}, and monetary policy \citep{Fan2010, Fan2012, Shang2023}. Additionally, when examining bond risk premiums, \citep{Bianchi2021, Zhai2024} have incorporated various macroeconomic variables, including consumption and taxation, with \cite{Jiang2024} analyzing 102 macroeconomic variables across 13 categories.

Our selection of macroeconomic variables is similar to that of \cite{Jiang2024}, utilizing 105 variables from 13 categories sourced from the Wind database. These categories include Macro-prosperity, Output, Consumption, Price Index, Interest Rates, Money and Credit, Investment, Real Estate, Tax, Trade, Foreign Exchange Rate, Stock Market, and Monetary Policy. Detailed variable names are provided in Appendix \ref{ufr:abbr_macro}.

\section{Empirical Analysis}

In this section, we primarily introduce the estimation results for UFRs, and the empirical results related to forecasting UFRs. We primarily employ the SDF, SFR and SYC methods, and the ZJW Improved method to derive endogenous UFRs, which serve as the main forecasting target. In the ZJW Improved method, the parameter $\lambda$, required for calibration, is derived using the de Kort-Vellekoop alternative methods, specifically the SFR and SYC methods. We incorporate the UFRs obtained from all four methods into our analysis and forecasting, rather than relying on any single method's UFR as the sole target. This approach ensures that the resulting analysis and predictions are more robust and reliable.

\subsection{Estimation Results for Ultimate Forward Rates} \label{est_for_ufr}
\begin{figure}[H]
	\centering
	\includegraphics[width=0.85\textwidth]{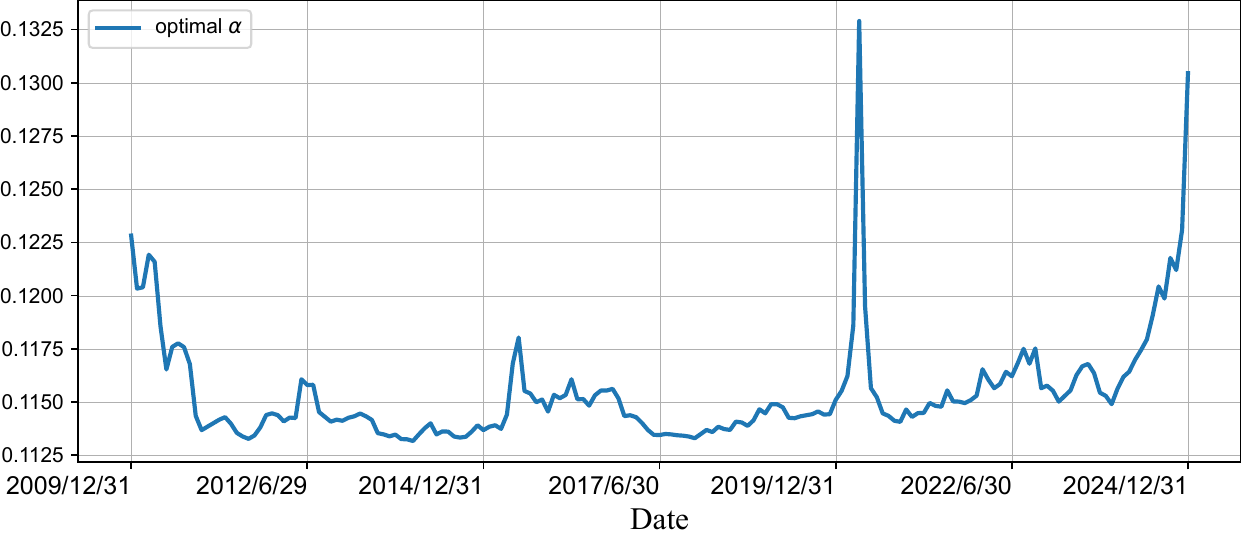}
	\caption{The optimal $\alpha$ of the ZJW Improved Method.\\ \footnotesize{Note: This figure illustrates the estimated dynamic $\alpha$, which exhibits unusual increases and volatility during periods of sharp declines in the UFR and ultra-long-term bond yields.}}  
	\label{ufr:alphas}
\end{figure}

\hyperref[alphas]{Fig. \ref{ufr:alphas}} presents the estimated optimal $\alpha$ for the ZJW Improved method. It is evident that during periods of extreme volatility in the bond market, such as the COVID-19 pandemic in early 2020 and the continued decline in China treasury bond yields at the end of 2024 (driven by investor expectations of low interest rates and deflation), $\alpha$ exhibits significant fluctuations. Notably, $\alpha$ is particularly sensitive to rapid declines in UFR; when the UFR declines sharply, $\alpha$ rises quickly.

\begin{figure}[H]
	\centering
	\includegraphics[width=0.9\textwidth]{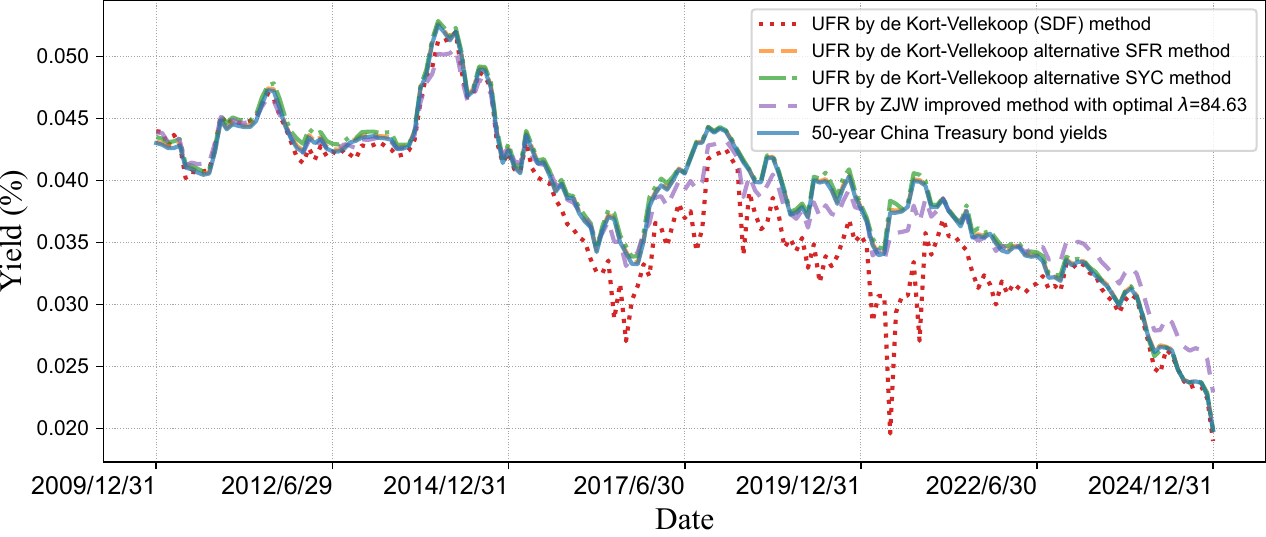}
	\caption{Estimation for UFRs.\\ \footnotesize{Note: This figure shows the UFR estimates obtained using different methods, along with the 50-year treasury bond yields. Overall, with the exception of the UFR estimated by the SDF method, the UFRs derived from the other methods align closely with the 50-year treasury bond yield curve. However, the UFR estimated by the SDF method diverges significantly after 2017, particularly exhibiting an abnormal decline and volatility around early 2020, which notably deviates from the other curves.}}  
	\label{ufr:UFRplot}
\end{figure}

\hyperref[UFRplot]{Fig. \ref{ufr:UFRplot}} presents the estimation results for the UFR, showing consistent trends across all four methods. However, the UFR derived from the SDF method exhibits abnormally low values at the start of 2020, reflecting its sensitivity to extreme market fluctuations, such as those induced by the COVID-19 pandemic. Additionally, from 2017 to 2023, the UFR derived from the SDF method shows significant volatility, which notably diverges from the results of the other methods and the actual 50-year treasury bond yields. These sharp and atypical fluctuations contradict the analysis in Section \ref{china_b_yields}, where the UFR is expected to remain relatively stable. In contrast, the ZJW Improved method, which builds upon the SDF method, mitigates these anomalies by incorporating prior information as a penalty term. Consequently, for the subsequent empirical analysis, we use the UFRs obtained from the SFR, SYC, and ZJW Improved methods as proxies for the UFR, discarding the UFR derived from the SDF method. Additionally, we incorporate the 30-year, 40-year, and 50-year ultra-long-term treasury bond yields for comparison.

\hyperref[table:unitroot]{Table \ref{ufr:table:unitroot}} presents the descriptive statistics and unit root tests for the UFR and ultra-long-term treasury bond yields. The results indicate that neither the UFR nor the ultra-long-term treasury bond yields series pass the ADF and PP unit root tests, suggesting that the series are non-stationary. After applying first-order differencing, both series pass the unit root tests. Given that non-stationary series can adversely affect the predictive performance of OLS and linear machine learning algorithms, we finally use the first-differenced series for the target variables, UFR and ultra-long-term treasury bond yields.

\begin{table}[H] 
	\caption{\\Descriptive Statistics and Unit Root Tests for UFR and Ultra-Long-Term Treasury Bond Yields}
	\setlength{\tabcolsep}{15pt}
	\renewcommand{\arraystretch}{0.8}
	\resizebox{1\textwidth}{!}{
		\begin{tabular}{ccccccc}
			\toprule
			& \textbf{Min}   & \textbf{Max}      & \textbf{Mean}   & \textbf{Std.}   & \textbf{ADF Statistic} & \textbf{PP Statistic} \\ \midrule
			$\mathbf{UFR_{SDF}}$             & 1.917  & 5.334 & 3.764  & 0.705 & -0.894        & -1.065       \\
			$\mathbf{UFR_{SFR}}$             & 2.012  & 5.407 & 3.982  & 0.636 & -0.109        & 0.316        \\
			$\mathbf{UFR_{SYC}}$             & 1.994  & 5.428 & 4.006  & 0.644 & -0.08         & 0.412        \\
			$\mathbf{UFR_{ZJW}}$             & 2.317  & 5.197 & 3.965  & 0.559 & 0.221         & 0.026        \\
			\textbf{30-YTM} & 1.912  & 5.114 & 3.786  & 0.6   & -0.069        & 0.201        \\
			\textbf{40-YTM}& 1.978  & 5.197 & 3.849  & 0.602 & -0.083        & 0.233        \\
			\textbf{50-YTM} & 1.988  & 5.252 & 3.892  & 0.611 & -0.078        & 0.338        \\
			$\mathbf{\Delta UFR_{SDF}}$      & -1.137 & 1.002 & -0.014 & 0.223 & -18.362***    & -18.715***   \\
			$\mathbf{\Delta UFR_{SFR}}$      & -0.384 & 0.371 & -0.013 & 0.12  & -10.033***    & -9.768***    \\
			$\mathbf{\Delta UFR_{SYC}}$      & -0.388 & 0.403 & -0.014 & 0.122 & -10.065***    & -9.760***    \\
			$\mathbf{\Delta UFR_{ZJW}}$      & -0.353 & 0.304 & -0.012 & 0.108 & -12.344***    & -12.389***   \\
			$\mathbf{\Delta}$ \textbf{30-YTM} & -0.358 & 0.349 & -0.013 & 0.115 & -10.271*** & -10.108*** \\
			$\mathbf{\Delta}$ \textbf{40-YTM} & -0.364 & 0.353 & -0.013 & 0.113 & -10.132*** & -9.948***  \\
			$\mathbf{\Delta}$ \textbf{50-YTM} & -0.363 & 0.349 & -0.013 & 0.115 & -9.995***  & -9.736***      \\ \bottomrule
		\end{tabular}
	}
	\begin{tablenotes}
		\fontsize{10}{10}\selectfont
		\item[1] Note: Augmented Dickey-Fuller test and Phillips-Perron test statistics with p-values below 0.1, 0.05, 0.01 are marked with ``*'',``**'',``***'', indicating significance at the 10\%, 5\%, 1\% level. 
	\end{tablenotes}
	\label{ufr:table:unitroot}
\end{table}

\subsection{Empirical Results of UFR Forecasting}
In this section, we address several key issues. First, we examine whether bond yields of different maturities have predictive power for the subsequent period's UFR. Second, we assess whether incorporating macroeconomic variables enhances the accuracy of the predictions. Third, we explore whether linear machine learning models outperform nonlinear models and investigate whether the relationships between the features used for prediction and the UFR are linear or nonlinear. Finally, we identify the features that contribute most to predicting the UFR.


In the study of the term structure of interest rates, \cite{DieboldLi2006} implemented two types of models within their competitive framework. One model directly regressed non-stationary bond yields for prediction, while the other used first differences of bond yields for regression-based forecasting. While machine learning models can often handle non-stationary time series, some models, such as OLS, may not be suitable for this task. As discussed in Section \ref{est_for_ufr} and presented in \hyperref[table:unitroot]{Table \ref{ufr:table:unitroot}}, both the UFR and ultra-long-term treasury bond yields fail the ADF and PP unit root tests, indicating they are non-stationary. After applying first-order differencing, both series pass the unit root tests. Given the negative impact non-stationary series can have on the predictive performance of OLS and linear machine learning algorithms, we apply first differencing to the UFR and ultra-long-term treasury bond yields for our forecasting models.

This study employs a forward rolling forecast approach. Specifically, we designate 75\% of the total sample as the sample window. Within each window, we estimate the model parameters and then predict the $\mathrm{\Delta UFR}$ for the subsequent period ($t+1$). For each forward rolling forecast, we re-estimate the model parameters and obtain a new prediction for $t+1$.

In particular, our total sample spans from January 29, 2010, to December 31, 2024, including monthly data on China treasury bond yields and macroeconomic factors, totaling 179 months\footnote{The starting time of our overall sample slightly differs from the data start time in Section \ref{china_b_yields} due to the use of first differences for both bond yields of different maturities and the UFR in the forecasting process.}. We set the window length to 134 months, resulting in an out-of-sample forecast period of 45 months.

\subsubsection{Model Evaluation Metrics}

We utilize the Root Mean Squared Error (RMSE) and Mean Absolute Error (MAE) to evaluate the prediction error. The RMSE and MAE are  defined as follows:
\begin{equation}
R M S E=\sqrt{\frac{1}{z} \sum_{i=1}^z\left(\mathrm{\Delta UFR}_{i}-\hat{\mathrm{\Delta UFR}}_{i}\right)^2}, \quad M A E=\frac{1}{n} \sum_{i=1}^n\left|\mathrm{\Delta UFR}_{i}-\hat{\mathrm{\Delta UFR}}_{i}\right|
\end{equation}
where $\hat{\mathrm{\Delta UFR}}_{i}$ and $\mathrm{\Delta UFR}_{i}$ represent the predicted and actual values of the UFR for the $i$-th observation, respectively, while $z$ denotes the total number of forecasts.


The forecasts generated by each model are compared to a naive benchmark, such as the historical mean of bond yields. To evaluate out-of-sample performance, the predictive $R^2$, as proposed by \cite{Campbell1991}, is adopted. The out-of-sample $R^2$ (denoted as $R_{\text{oos}}^2$) is defined as:

\begin{equation}
R_{\text {oos }}^2=1-\frac{\sum_{t_0=h}^{T-h}\left(y_{t+1}^{(n)}-\widehat{y}_{t+1}^{(n)}\left(\mathcal{M}_c\right)\right)^2}{\sum_{t_0=h}^{T-h}\left(y_{t+1}^{(n)}-\widehat{y}_{t+1}^{(n)}\left(\mathcal{M}_{b}\right)\right)^2},
\end{equation}
where $T$ denotes the length of the out-of-sample period, and $\widehat{y}_{t+1}^{(n)}\left(\mathcal{M}_{c}\right)$ denotes the one-month-ahead forecast of bond yields for maturity $n$ generated using complex model $\mathcal{M}_c$, while $\widehat{y}_{t+1}^{(n)}\left(\mathcal{M}_{b}\right)$ represents the one-month-ahead prediction error based on the benchmark model $\mathcal{M}_{b}$. 

Additionally, we utilize the modified MSE introduced by \cite{Clark2007}:
\begin{equation}
d^{C W}_{t, \tau_n} = \left(y_{t, \tau_n}-\hat{y}_{t  , \tau_n}^{\mathcal{M}_{c}}\right)^2 - \left(y_{t, \tau_n}-\hat{y}_{t, \tau_n}^{\mathcal{M}_{b}}\right)^2  + \left(\hat{y}_{t, \tau_n}^{\mathcal{M}_{b}} - \hat{y}_{t, \tau_n}^{\mathcal{M}_{c}}\right)^2.
\end{equation}
The Clark and West test is employed to determine whether the performance of the more complex model exceeds that of the benchmark model. Statistical significance from the Clark and West test will be reported if $R_{O O S}^2\left(\mathcal{M}_{c}, \mathcal{M}_{b}\right)>0$.

\subsubsection{Forecasting UFR with Bond Yields}

In the calculation of $R_{\text{oos}}^2$, we select the random walk model as the benchmark, which is also used as a benchmark model in \cite{DieboldLi2006} and \cite{Fernandes2019}. The primary reason for this choice is to test whether our model, $\mathcal{M}_c$, demonstrates predictive capability. 

Panels A, B, and C in \hyperref[table:oosr2_only_yields]{Table \ref{table:oosr2_only_yields}} display the $R_{\text{oos}}^2$ for the OLS model, linear machine learning models, and nonlinear machine learning models, respectively, using bond yields as the sole features for rolling predictions. The target time series for prediction are the first-order differences of the UFRs obtained through three different methods, as well as the first-order differences of three ultra-long-term bond yields. 

The performance of the OLS model is suboptimal, with $R_{\text{oos}}^2$ hovering around zero, particularly in the prediction of $\operatorname{\Delta UFR_{SFR}}$, $\operatorname{\Delta UFR_{SYC}}$, and $\Delta$ 50-YTM. After incorporating penalty terms, the performance of Lasso, Ridge, and Elastic Net improves, with significant positive $R_{\text{oos}}^2$ observed for the predictions of $\operatorname{\Delta UFR_{SFR}}$, $\operatorname{\Delta UFR_{SYC}}$, $\operatorname{\Delta UFR_{ZJW}}$, and $\Delta$ 30-YTM.

\begin{table}[H]
	\caption{\\$R_{\text{oos}}^2$: Forward Rolling Forecast with Bond Yields  vs. Random Walk}
	\setlength{\tabcolsep}{5pt}
	\renewcommand{\arraystretch}{1}
	\resizebox{1\textwidth}{!}{
	\begin{tabular}{lcccccc}
	\toprule
	\textbf{} & $\mathbf{\Delta UFR_{SFR}}$ & $\mathbf{\Delta UFR_{SYC}}$ & $\mathbf{\Delta UFR_{ZJW}}$ & $\mathbf{\Delta}$ \textbf{30-YTM} & $\mathbf{\Delta}$ \textbf{40-YTM} & $\mathbf{\Delta}$ \textbf{50-YTM} \\ \midrule
	\textbf{Panel   A: OLS}                                &           &           &           &           &           &           \\ \hline
	\textbf{OLS}                                           & -3.63\%   & -5.33\%   & 1.80\%*** & 4.30\%*** & 1.32\%*** & -2.70\%    \\ \hline
	\textbf{Panel   B: Linear machine learning methods}    &           &           &           &           &           &           \\ \hline
	\textbf{Penalized   linear regressions}                &           &           &           &           &           &           \\ 
	\textbf{Lasso}                                         & 6.61\%*** & 7.68\%*** & 8.16\%*** & 3.04\%*** & -0.11\%   & 2.44\%*** \\ 
	\textbf{Ridge}                                         & 6.29\%*** & 7.43\%*** & 9.07\%*** & 3.35\%*** & -3.33\%   & -1.54\%   \\ 
	\textbf{Elastic net}                                   & 5.88\%*** & 7.20\%*** & 8.66\%*** & 1.83\%*** & 2.74\%*** & -0.18\%   \\ 
	\textbf{PCR and PLS}                                 &           &           &           &           &           &           \\ 
	\textbf{PCR (3 components)}                            & 2.26\%*** & 4.07\%*** & 8.54\%*** & -0.46\%   & -2.32\%   & -0.02\%   \\ 
	\textbf{PCR (5 components)}                            & -1.92\%   & -0.25\%   & 3.33\%*** & -2.10\%   & -0.98\%   & -0.52\%   \\ 
	\textbf{PCR (10 components)}                           & -0.78\%   & -2.91\%   & 2.25\%*** & 3.83\%*** & 2.58\%*** & -2.29\%   \\ 
	\textbf{PLS (3 components)}                            & -0.16\%   & 1.54\%*** & 4.12\%*** & -2.28\%   & 1.26\%    & 3.13\%    \\ 
	\textbf{PLS (5 components)}                            & -0.74\%   & -1.00\%   & 6.44\%*** & -2.90\%   & 4.79\%    & 3.97\%    \\ 
	\textbf{PLS (10 components)}                           & -3.21\%   & -5.71\%   & 1.38\%*** & 4.66\%*** & 7.90\%*** & 9.72\%    \\ 
	\hline
	\textbf{Panel   C: Nonlinear machine learning methods} &           &           &           &           &           &           \\ \hline
	\textbf{Regression   trees}                            &           &           &           &           &           &           \\ 
	\textbf{Random forest}                                 & 2.96\%*** & 3.92\%*** & 2.50\%*** & 1.00\%*** & 7.62\%*** & 8.34\%*** \\ 
	\textbf{Gradient-boosted trees}                        & 3.83\%*** & 4.82\%*** & 6.10\%*** & 4.57\%*** & 5.39\%*** & 6.42\%*** \\ 
	\textbf{XGBoost}                                       & 9.20\%*** & 5.71\%*** & 11.76\%*** & 7.70\%*** & 7.77\%*** & 7.85\%*** \\ 
	\textbf{Neural   networks}                             &           &           &           &           &           &           \\ 
	\textbf{Yield-Only-Net 1 layer (3 nodes)}                          & 13.53\%*** & 11.80\%*** & \textbf{16.00\%***} & 10.45\%*** & 14.61\%*** & 13.44\%*** \\ 
	\textbf{Yield-Only-Net 1 layer (5 nodes)}                          & 7.38\%*** & 6.96\%*** & 2.85\%*** & 3.26\%*** & 3.90\%*** & 7.64\%*** \\ 
	\textbf{Yield-Only-Net 1 layer (7 nodes)}                          & 9.40\%*** & 10.90\%*** & 9.46\%*** & 6.85\%*** & 8.34\%*** & 11.20\%***  \\ 
	\textbf{Yield-Only-Net 2 layer (3 nodes)}                          & 7.15\%*** & 6.85\%*** & 4.70\%*** & 2.76\%*** & 10.40\%*** & 11.59\%*** \\ 
	\textbf{Yield-Only-Net 2 layer (5 nodes)}                          & 7.94\%*** & 7.40\%*** & 11.30\%*** & 4.39\%*** & 10.48\%*** & 11.71\%*** \\ 
	\textbf{Yield-Only-Net 2 layer (7 nodes)}                          & 9.24\%*** & 8.91\%*** & 10.06\%*** & 6.53\%*** & \textbf{15.66\%***} & \textbf{15.53\%***} \\ 
	\textbf{Yield-Only-Net 3 layer (3 nodes)}                          & 10.17\%*** & 8.84\%*** & 8.74\%*** & 6.72\%*** & 4.64\%*** & 6.70\%***  \\ 
	\textbf{Yield-Only-Net 3 layer (5 nodes)}                          & 12.22\%*** & 10.21\%*** & 10.81\%*** & 10.25\%*** & 4.18\%*** & 6.30\%***  \\ 
	\textbf{Yield-Only-Net 3 layer (7 nodes)}                          & \textbf{15.12\%***} & \textbf{15.03\%***} & 7.80\%***  & \textbf{13.12\%***} & 3.23\%*** & 5.93\%***  \\ \bottomrule
	\end{tabular}
	}
	\begin{tablenotes}
		\fontsize{10}{10}\selectfont
		\item[1] Note: Subscripts SFR, SYC, and ZJW denote UFRs derived from the Smoothest Forward Rate, Smoothest Yield Curve, and ZJW Improved methods, respectively. Clark and West (CW) statistics are annotated with ``*'', ``**'', and ``***'' to indicate significance at the 10\%, 5\%, and 1\% levels. Significance levels are reported only when $R_{\text{oos}}^2 > 0$. 
	\end{tablenotes}
	\label{table:oosr2_only_yields}
\end{table}

The PCR and PLS models, regardless of whether 3, 5, or 10 components are used, fail to demonstrate satisfactory performance, with $R_{\text{oos}}^2$ consistently close to zero. However, all models show a significant positive $R_{\text{oos}}^2$ for $\operatorname{\Delta UFR_{ZJW}}$, indicating that the first-order difference of bond yields carries predictive power for $\operatorname{\Delta UFR_{ZJW}}$. 

Panel C of \hyperref[table:oosr2_only_yields]{Table \ref{table:oosr2_only_yields}} shows the $R_{\text{oos}}^2$ for the rolling predictions of nonlinear models. Overall, both regression trees and neural network models show positive $R_{\text{oos}}^2$, suggesting a nonlinear relationship between the first-order difference of bond yields and the $\operatorname{\Delta UFR}_{t+1}$ as well as $\Delta$-ultra-long-term bond yield. Specifically, in the regression tree models, XGBoost performs slightly better, though the difference is not substantial. In contrast, neural network models demonstrate strong overall predictive performance. The Yield-Only-Net 1-layer (3 nodes) model achieves over 10\% $R_{\text{oos}}^2$ for all time series, while the Yield-Only-Net 3-layer (7 nodes) and Yield-Only-Net 2-layer (7 nodes) models perform best in all predictions except for $\operatorname{\Delta UFR_{ZJW}}$, highlighting the superior predictive capability of neural networks. Additionally, we find no significant improvement in prediction with more layers or nodes in the neural network models.

\subsubsection{Incorporating Macroeconomic Variables in UFR Prediction}

Subsequently, we incorporate macroeconomic variables into the analysis, meaning that the features used for prediction now include both bond yields and macroeconomic variables. The results of the rolling predictions using features that include macroeconomic variables are presented in \hyperref[table:oosr2_macro_yields]{Fig. \ref{table:oosr2_macro_yields}}. The key changes in these models lie in the structures of the neural network models.

Panel A of \hyperref[table:oosr2_macro_yields]{Table \ref{table:oosr2_macro_yields}} presents the rolling forecast results of the OLS model. Due to its inherent limitations in handling a large number of macroeconomic variables, we apply principal component analysis (PCA) to extract principal components from 105 macroeconomic variables, categorized into 13 groups: Macro-prosperity, Output, Consumption, Price Index, Interest Rates, Money and Credit, Investment, Real Estate, Tax, Trade, Foreign Exchange Rate, Stock Market, and Monetary Policy. 

\begin{table}[H]
	\caption{\\$R_{\text{oos}}^2$: Forward Rolling Forecast with Bond Yields and Macroeconomic Variables vs. Random Walk}
	\setlength{\tabcolsep}{3pt}
	\renewcommand{\arraystretch}{1}
	\resizebox{\textwidth}{!}{
	\begin{tabular}{lcccccc}
	\toprule
	\textbf{} & $\mathbf{\Delta UFR_{SFR}}$ & $\mathbf{\Delta UFR_{SYC}}$ & $\mathbf{\Delta UFR_{ZJW}}$ & $\mathbf{\Delta}$ \textbf{30-YTM} & $\mathbf{\Delta}$ \textbf{40-YTM} & $\mathbf{\Delta}$ \textbf{50-YTM} \\ \midrule
	\textbf{Panel   A: OLS}                                &           &           &           &           &           &           \\ \hline
	\textbf{OLS}                                           & -45.6\%   & -49.8\%   & -34.14\%  & -25.3\%   & -33.63\%  & -43.13\%  \\ \hline
	\textbf{Panel   B: Linear machine learning methods}    &           &           &           &           &           &           \\ \hline
	\textbf{Penalized   linear regressions}                &           &           &           &           &           &           \\
	\textbf{Lasso}                                         & 10.1\%***  & 11\%***   & 10.82\%*** & 3.31\%***  & 7.18\%***  & 9.84\%***  \\
	\textbf{Ridge}                                         & 11.55\%*** & 12.16\%*** & 17.28\%*** & 7.34\%***  & 9.31\%***  & 11.85\%*** \\
	\textbf{Elastic net}                                   & 8.58\%***  & 10.76\%*** & 12.41\%*** & 2.19\%***  & 4.62\%***  & 8.96\%***  \\
	\textbf{PCR and   PLS}                                 &           &           &           &           &           &           \\
	\textbf{PCR (3 components)}                            & 15.69\%*** & 16.9\%***  & 15.32\%*** & 7.41\%***  & 11.8\%***  & 15.56\%*** \\
	\textbf{PCR (5 components)}                            & 14.55\%*** & 16.44\%*** & 16.87\%*** & 6.72\%***  & 10.09\%*** & 14.5\%***  \\
	\textbf{PCR (10 components)}                           & 5.44\%***  & 8.33\%***  & 12.77\%*** & -1.13\%   & 0.72\%***  & 5.81\%***  \\
	\textbf{PLS (3 components)}                            & -13.35\%  & -9.16\%   & 4.71\%***  & -7.88\%   & -14.39\%  & -12.23\%  \\
	\textbf{PLS (5 components)}                            & -31.48\%  & -25.35\%  & -8.79\%   & -19.63\%  & -31.48\%  & -29.77\%  \\
	\textbf{PLS (10 components)}                           & -168.93\% & -157.6\%  & -133.97\% & -131.12\% & -160.41\% & -164.08\% \\ \hline
	\textbf{Panel   C: Nonlinear machine learning methods} &           &           &           &           &           &           \\ \hline
	\textbf{Regression   trees}                            &           &           &           &           &           &           \\
	\textbf{Random forest}                                 & 11.75\%*** & 9.88\%***  & 10.75\%*** & 4.37\%***  & 9.05\%***  & 12.4\%***  \\
	\textbf{Gradient-boosted trees}                        & 11.04\%*** & 10.09\%*** & 10.87\%*** & 5.39\%***  & 9.7\%***   & 10.9\%***  \\
	\textbf{XGBoost}                                       & 11.32\%*** & 11.05\%*** & 8.53\%***  & 3.74\%***  & 9.06\%***  & 11.38\%*** \\
	\textbf{Neural networks}                             &           &           &           &           &           &           \\
	\textbf{Yield-Macro-Net 1 layer (32 nodes)}               & 13.96\%*** & 13.93\%*** & 19.08\%*** & 10.75\%*** & 12.54\%*** & 14.13\%*** \\
	\textbf{Yield-Macro-Net 2 layers (32, 16 nodes)}          & 16.75\%*** & 16.69\%*** & \textbf{21.05\%***} & 14.8\%***  & 16.37\%*** & 17.16\%*** \\
	\textbf{Yield-Macro-Net 3 layers (32, 16, 8 nodes)}       & \textbf{16.76\%***} & 15.79\%*** & 15.49\%*** & \textbf{18.51\%***} & \textbf{18.59\%***} & 18.44\%*** \\
	\textbf{Hybrid net 1   layer (32)}         & 14.08\%*** & \textbf{22.18\%***} & 8.96\%***  & 11.54\%*** & 15.99\%*** & \textbf{18.82\%***} \\
	\textbf{Hybrid net 2   layers (32, 16)}    & 12.4\%***  & 15.34\%*** & 18.79\%*** & 9.16\%***  & 9.13\%***  & 11.79\%*** \\
	\textbf{Hybrid net 3   layers (32, 16, 8)} & 13.75\%*** & 11\%***   & 13.9\%***  & 6.58\%***  & 9.79\%***  & 9.37\%***  \\
	\textbf{Double net 1   layer}              & 15.94\%*** & 16.09\%*** & 17.92\%*** & 10.84\%*** & 13.7\%***  & 15.78\%*** \\
	\textbf{Double net 2   layers}             & 15.27\%*** & 14.35\%*** & 8.47\%***  & 8.6\%***   & 12.9\%***  & 15.47\%*** \\
	\textbf{Double net 3   layers}             & 11.9\%***  & 10.52\%*** & 16.66\%*** & 11.12\%*** & 12.59\%*** & 11.89\%*** \\
	\textbf{GN-Net  1 layer (1 node per group, 3 nodes)} & 16.11\%*** & 17.66\%*** & 19.52\%*** & 11.08\%*** & 11.89\%*** & 15.58\%*** \\
	\textbf{GN-Net  2 layers (2 nodes per group, 3 nodes)} & 12.49\%*** & 11.44\%*** & 7.1\%***   & 10.12\%*** & 11.13\%*** & 12.05\%*** \\
	\textbf{GN-Net 3 layers (3 nodes per group, 3 nodes)} & 13.89\%*** & 13\%***   & 10.63\%*** & 12.4\%***  & 14.41\%*** & 15.22\%*** \\ \bottomrule
	\end{tabular}
	}
	\begin{tablenotes}
		\fontsize{10}{10}\selectfont
		\item[1] Note: Subscripts SFR, SYC, and ZJW denote UFRs derived from the Smoothest Forward Rate, Smoothest Yield Curve, and ZJW Improved methods, respectively. CW statistics are annotated with ``*'', ``**'', and ``***'' to indicate significance at the 10\%, 5\%, and 1\% levels. Significance levels are reported only when $R_{\text{oos}}^2 > 0$. 
	\end{tablenotes}
	\label{table:oosr2_macro_yields}
\end{table}

Despite this dimensionality reduction, the OLS model performs poorly, with all $R_{\text{oos}}^2$ values being large negative numbers. This indicates that incorporating the macroeconomic principal components worsens the predictions, yielding worse results than using only the first differences of bond yields as input features. In contrast, penalized models such as Lasso, Ridge, and Elastic Net exhibit significant improvements in prediction performance. For all time series, the $R_{\text{oos}}^2$ values calculated for these models are significantly positive, with the Ridge model, using L2 regularization, performing the best.

In linear models, the performance of PCR and PLS is notably divergent. PCR performs relatively well, especially when using the first three or five principal components, where the $R_{\text{oos}}^2$ values are significantly positive. However, increasing the number of principal components beyond ten leads to a significant deterioration in PCR’s performance, with a negative $R_{\text{oos}}^2$ value observed in the prediction of $\Delta 30$-YTM. On the other hand, PLS performs very poorly. Regardless of whether the first three, five, or ten principal components are used, its performance remains unsatisfactory, with almost all $R_{\text{oos}}^2$ values negative. Overall, adding macroeconomic variables as input features improves the predictive performance of the PCR and penalized linear regression models, while significantly reducing the predictive accuracy of the OLS and PLS models.

In nonlinear models, the performance of regression tree-based models improves with the inclusion of macroeconomic variables as input features, particularly for Random Forest and Gradient-Boosted Trees. Both models exhibit $R_{\text{oos}}^2$ values exceeding 10\% in the prediction of $\operatorname{\Delta UFR_{SFR}}$, $\operatorname{\Delta UFR_{ZJW}}$, and $\Delta 50$-YTM. Similarly, neural network models show comparable improvements, with the Yield-Macro-Net 3-layer model (32, 16, 8 nodes) achieving $R_{\text{oos}}^2$ values over 15\% in all predictions. Additionally, the Hybrid Net 1-layer (32 nodes) and Yield-Macro-Net 2-layer (32, 16 nodes) models achieve $R_{\text{oos}}^2$ values exceeding 20\% in the predictions of $\operatorname{\Delta UFR_{SYC}}$ and $\operatorname{\Delta UFR_{ZJW}}$, respectively. However, the predictive performance of neural network models does not exhibit significant improvements as the number of layers, nodes, or model complexity increases. Overall, neural networks demonstrate the best performance, with more stable and superior results when macroeconomic variables are included as input features, compared to using only the first differences of bond yields. The superior performance of nonlinear models highlights the nonlinear relationship between input features and $\operatorname{\Delta UFR}$ as well as $\Delta$-ultra-long-term bond yields.

\subsubsection{Relative Importance of Macroeconomic Variables}

In this section, we select the Yield-Macro-Net 3-layer model (32, 16, 8 nodes), which performs best in predictions, as the forecasting model. Following the structure outlined in \cite{Zhai2024}, we employ SHapley Additive exPlanations (SHAP) to analyze the absolute SHAP values of individual macroeconomic variables, ranking them to explore the importance of each variable. Additionally, we rank the group importance by summing the SHAP values  and absolute SHAP values of the variables within each group, thereby investigating the significance of macroeconomic categories in the prediction process.

\begin{figure}[H]
	\centering
	\subfigure[$\mathrm{\Delta UFR_{ZJW}}$.]{
		\includegraphics[width=0.48\textwidth]{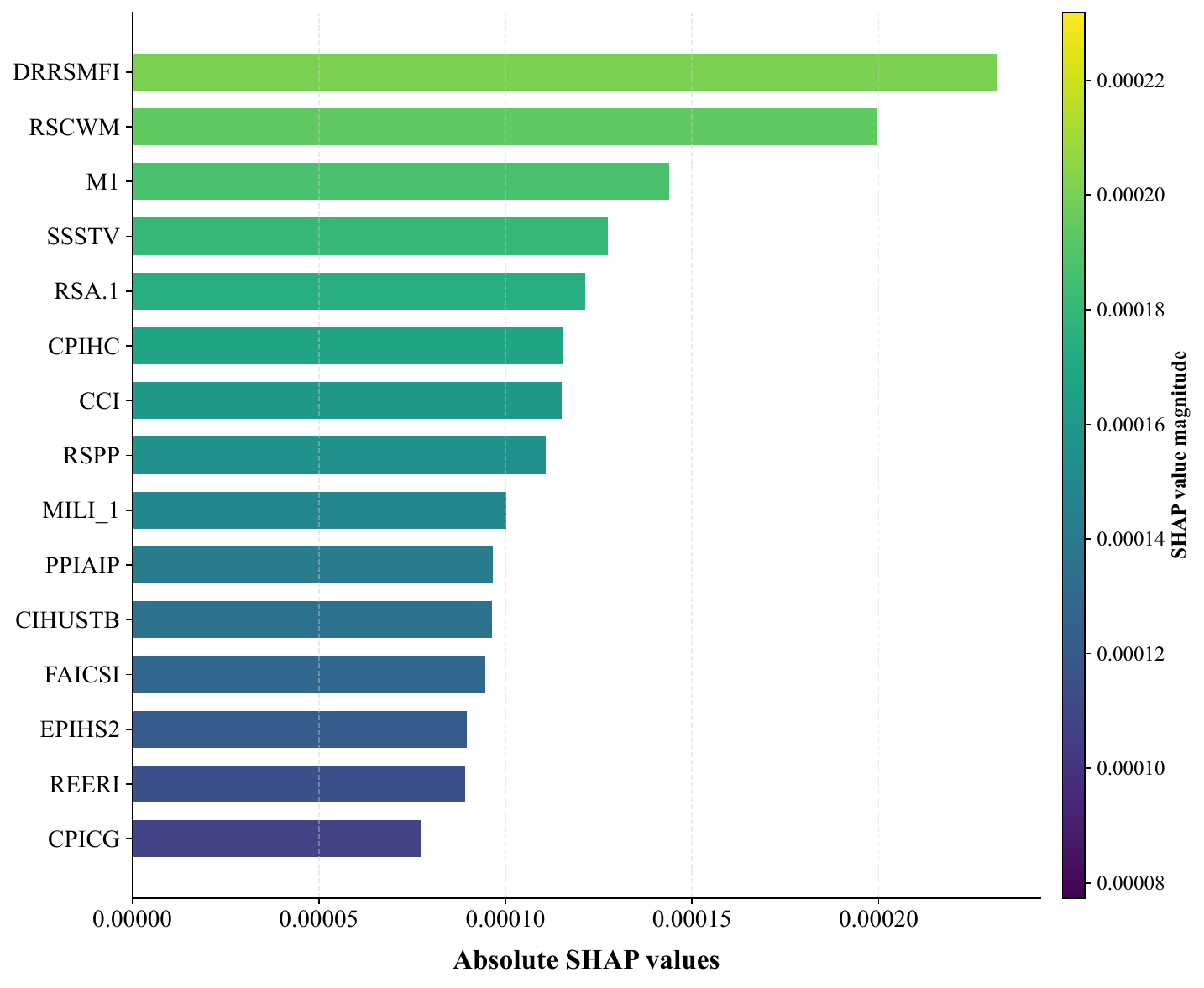}
	}
	\hspace{-3mm}
	\subfigure[$\mathrm{\Delta}$ 50-year YTM.]{
		\includegraphics[width=0.48\textwidth]{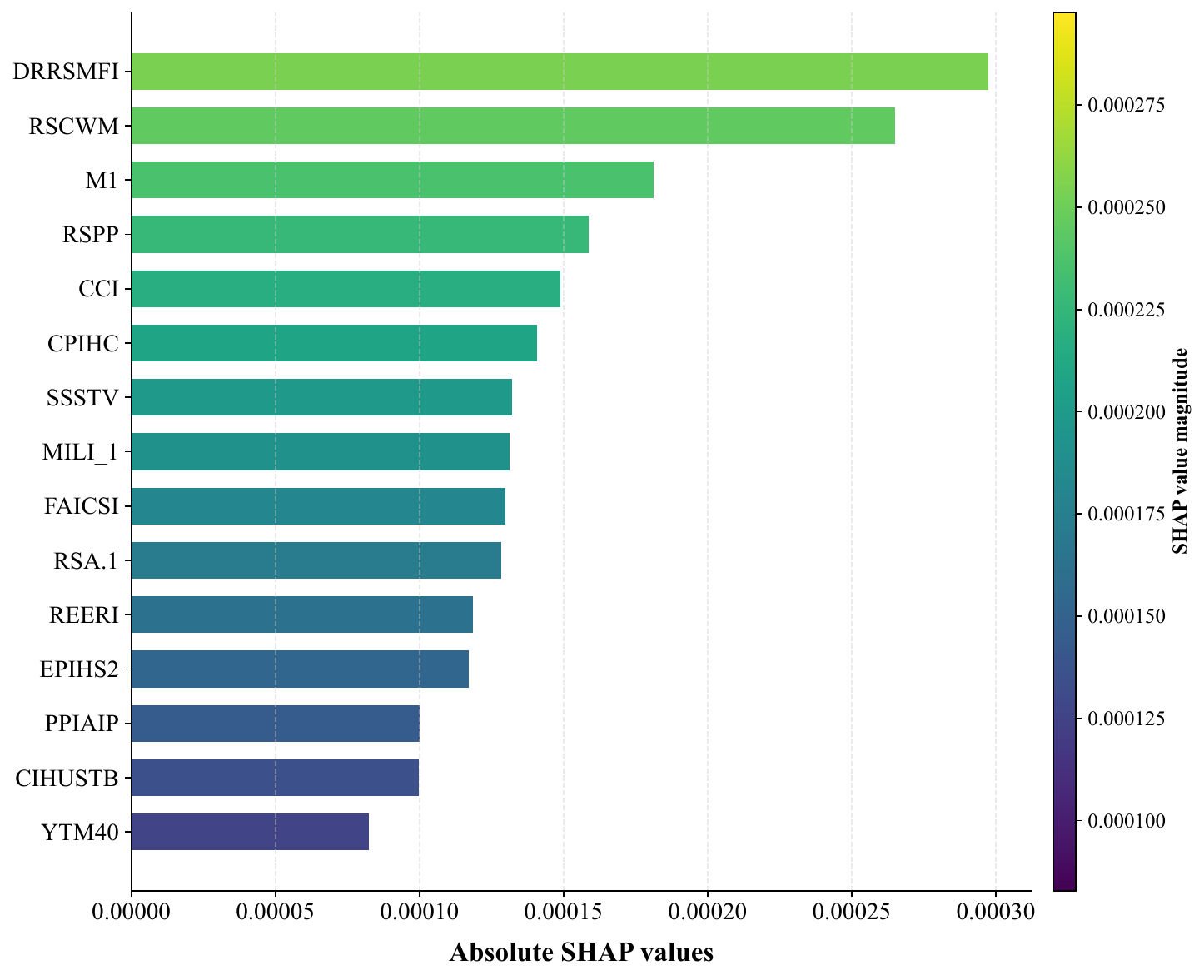}
	}
	\caption{Absolute SHAP value: individual variables.\\ \footnotesize{Note: This figure illustrates the absolute SHAP values of individual variables. We use the Yield-Macro-Net 3-layer model (32, 16, 8 nodes) to predict $\Delta$ UFR and $\Delta$-ultra-long-term bond yields. For $\Delta$ UFR, we select the representative variable $\operatorname{\Delta UFR_{ZJW}}$, while for $\Delta$-ultra-long-term bond yields, we select the longest maturity variable, $\Delta 50$-YTM.}}  
	\label{individual_abs_shap}
\end{figure}

%

\begin{figure}[H]
	\subfigure[$\mathrm{\Delta UFR_{ZJW}}$.]{
		\includegraphics[width=0.48\textwidth]{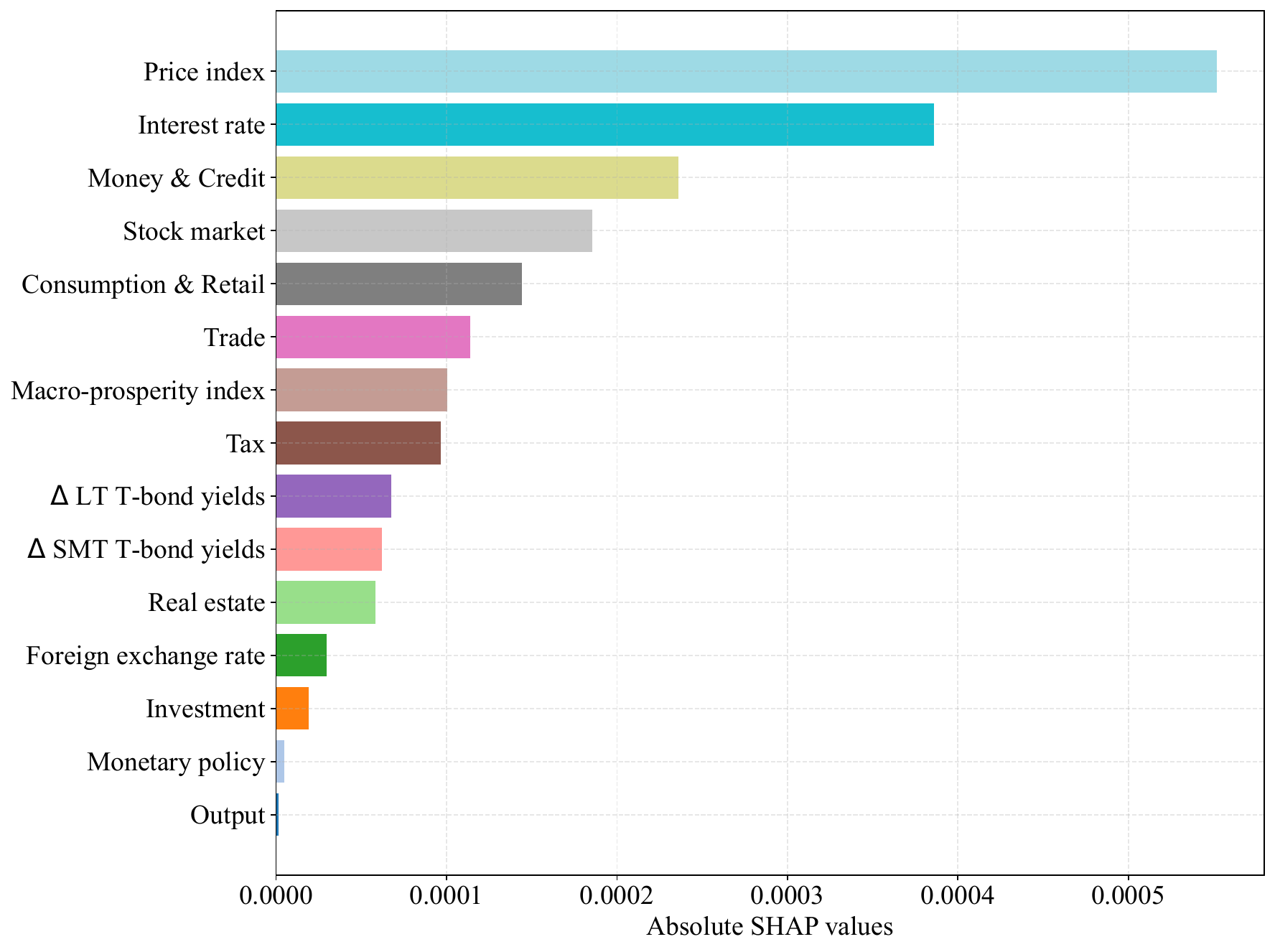}
	}
	\hspace{-3mm}
	\subfigure[$\mathrm{\Delta}$ 50-year YTM.]{
		\includegraphics[width=0.48\textwidth]{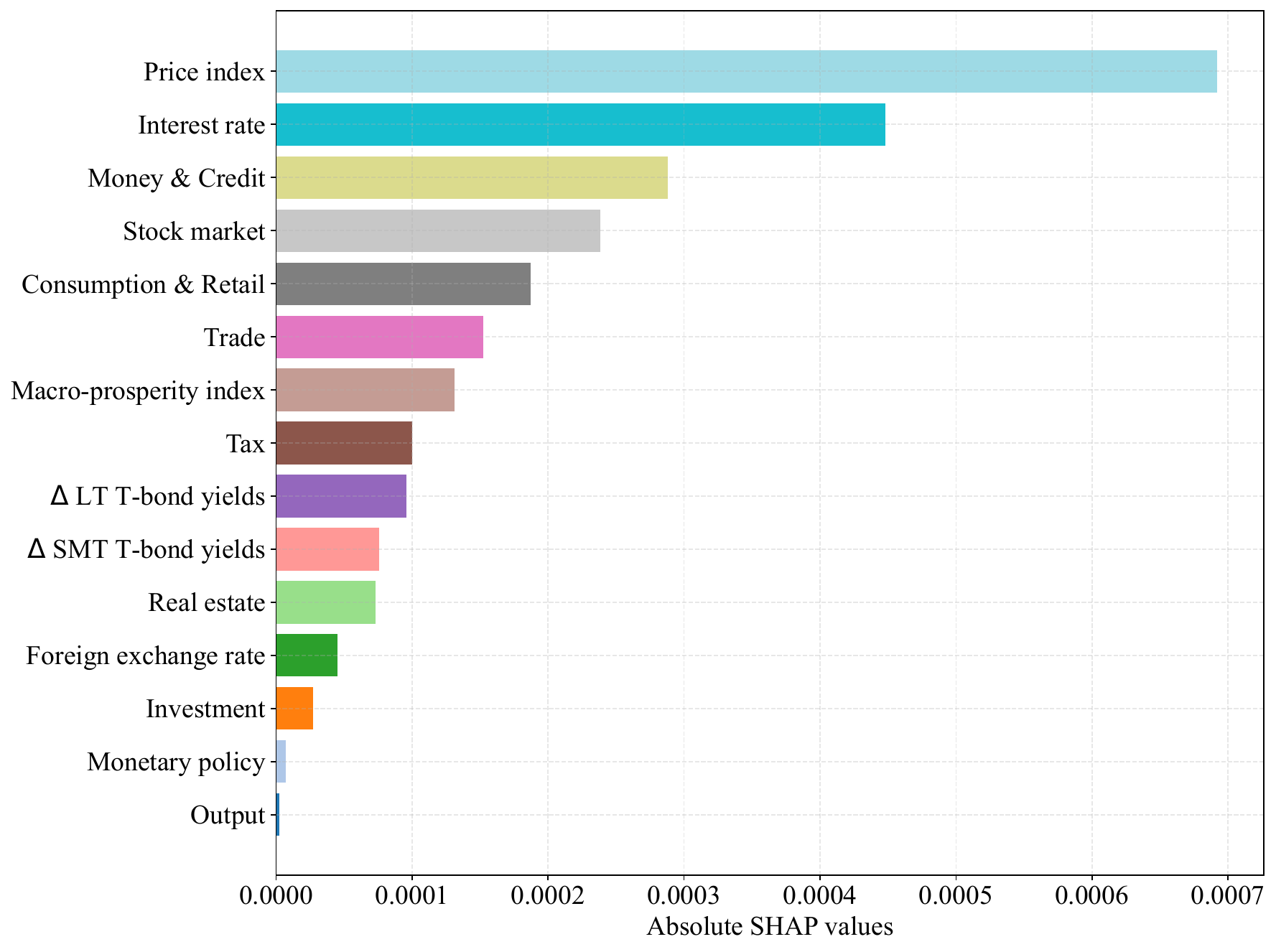}
	}
	\caption{Absolute SHAP value: group‐level.\\ \footnotesize{Note: This figure shows the group-level absolute SHAP values, where the absolute SHAP values of individual variables are summed by group and then ranked. The model used is the best-performing Yield-Macro-Net 3-layer model (32, 16, 8 nodes), and the figure presents the results of this model in predicting $\operatorname{\Delta UFR_{ZJW}}$ and $\Delta 50$-YTM.}}  
	\label{cate_abs_shap}
\end{figure}

\hyperref[individual_abs_shap]{Fig. \ref{individual_abs_shap}} and \hyperref[cate_abs_shap]{Fig. \ref{cate_abs_shap}} present the absolute SHAP values of individual macroeconomic variables and macroeconomic variable groups, respectively. From these figures, we observe that the RMB Deposit Reserve Ratio for Small and Medium-Sized Deposit Financial Institutions (RDRRSMDFI) has the greatest impact on the predictions of $\operatorname{\Delta UFR_{ZJW}}$ and $\Delta 50$-YTM. This is logical, as changes in the deposit reserve ratio affect liquidity levels in the market. 

\hyperref[cate_abs_shap]{Fig. \ref{cate_abs_shap}} further aggregates the absolute SHAP values by macroeconomic variable group, revealing that the Price Index, Interest Rate, and Money and Credit categories rank as the top three. In particular, the Price Index has the largest impact on the predictions of $\operatorname{\Delta UFR_{ZJW}}$ and $\Delta 50$-YTM. This finding is consistent with prior research, where Price Index-related macroeconomic variables played a significant role in bond yield modeling and forecasting, as demonstrated in \cite{Diebold2006} and \cite{Fernandes2019}. Similarly, Interest Rate and Money and Credit are typically closely linked to bond yield fluctuations. 

\begin{figure}[H]
	\subfigure[$\mathrm{\Delta UFR_{ZJW}}$.]{
		\includegraphics[width=0.48\textwidth]{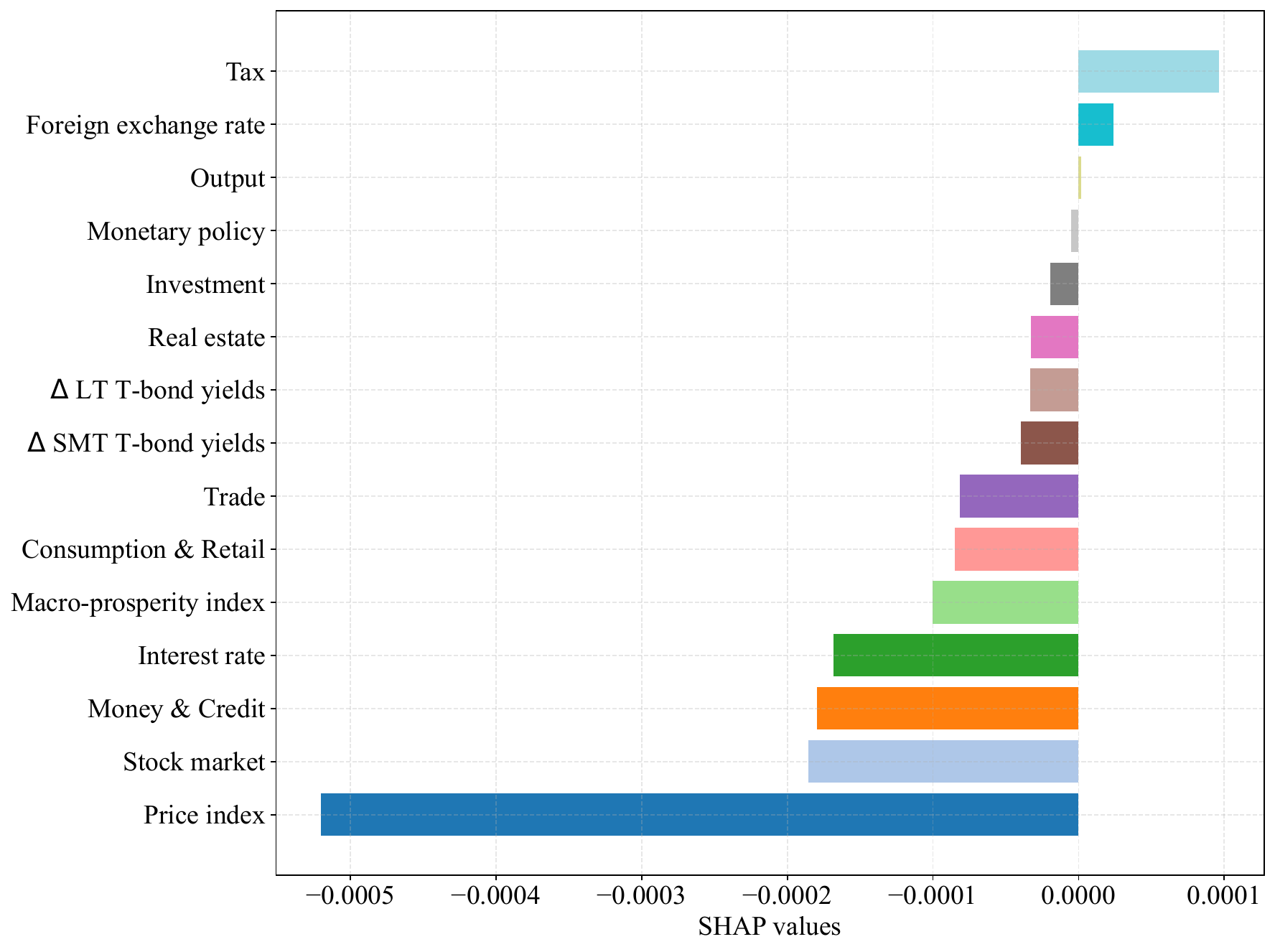}
	}
	\hspace{-3mm}
	\subfigure[$\mathrm{\Delta}$ 50-year YTM.]{
		\includegraphics[width=0.48\textwidth]{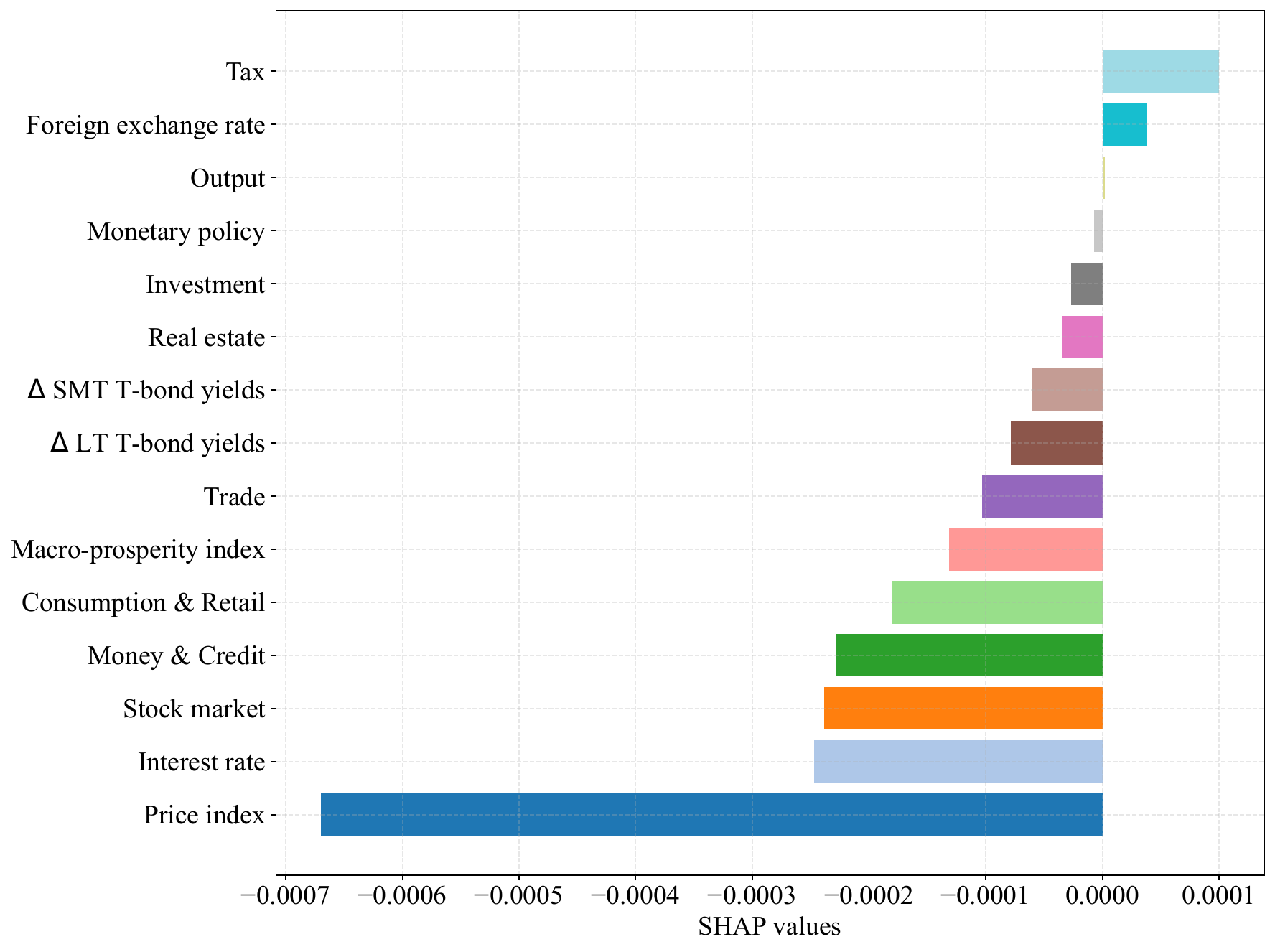}
	}
	\caption{Shap value: group‐level.\\ \footnotesize{Note: This figure shows the group-level SHAP values, where the SHAP values of individual variables are summed by group and then ranked. The sign of the SHAP values is retained to reflect the direction of influence of the macroeconomic variable groups. The model used is the best-performing Yield-Macro-Net 3-layer model (32, 16, 8 nodes), and the figure presents the results of this model in predicting $\operatorname{\Delta UFR_{ZJW}}$ and $\Delta 50$-YTM.}}
	\label{cate_shap}
\end{figure}

\hyperref[cate_shap]{Fig. \ref{cate_shap}} further illustrates the signs of the SHAP values, aggregated by macroeconomic variable group. The negative SHAP value for the Price Index indicates that fluctuations in the price index contribute to the decline in both ultra-long-term bond yields and UFR. In comparison, the impact of other macroeconomic categories is less significant than that of the Price Index. 

In summary, the Price Index plays a significant role in the prediction of $\operatorname{\Delta UFR_{ZJW}}$ and $\Delta 50$-YTM, which may partly explain why the inclusion of macroeconomic variables leads to a substantial improvement in prediction performance when using nonlinear machine learning models.

\section{UFR-Based Bond Yields Prediction}
In this section, we aim to develop a forecasting model for bond yields at different maturities based on the predicted UFR. As demonstrated in previous sections, we utilize various machine learning algorithms to forecast $\operatorname{\Delta UFR}$. Using the predicted $\operatorname{\Delta UFR}$, the forecast for $\widehat {\operatorname{UFR}}_{t+1}$ can be expressed as follows:

\begin{equation} \label{ufr_forecasts}
\begin{aligned}
\widehat {\operatorname{UFR}}_{t+1}= \operatorname{UFR}_{t} + \widehat {\operatorname{\Delta UFR}}_{t+1}
\end{aligned}
\end{equation}

For zero-coupon treasury bonds, the coupons $c_{i, j}$ in \hyperref[SW-def]{Eq. (\ref{SW-def})} becomes the identity matrix. Thus, the expression for bond prices is as follows:

\begin{equation} 
\begin{aligned}
m_i
&=e^{-f_{\infty} \tau} + \sum_{i=1}^N \xi_i \sum_{j=1}^J W\left(\tau, u_j\right),
\end{aligned}
\end{equation} 
and in matrix form, this can be expressed as:

\begin{equation}
\begin{aligned}
\mathbf{m}_{t}=&\mathbf{P}_{t}^{\top}\\
=&\mathbf{q}_{t} + \bm{\xi}_{t} \mathbf{W}_{t}, \quad t=1,\ldots, T.
\end{aligned}
\end{equation}

To forecast the bond prices for various maturities at time $t+1$, we employ the Smith-Wilson method combined with the predicted $\widehat {\operatorname{UFR}}_{t+1}$. Specifically, the parameter matrix $\boldsymbol{\xi}_t$, which is obtained by fitting bond yields using the Smith-Wilson method at time $t$, will be used in the forecast for $t+1$, while $\boldsymbol{\xi}_t$ remains unchanged. In the prediction for $t+1$, $\widehat {\mathbf{q}}_t$ and $\widehat {\mathbf{W}}_{t+1}$ are determined by UFR, and their estimated values can be expressed as follows:

\begin{equation} \label{q_forecsts}
\begin{aligned}
\widehat {\mathbf{q}}_{t+1}&=e^{-\widehat {f}_{\infty,t+1} \tau}\\
&=e^{- \ln(1+ \widehat {\operatorname{UFR}}_{t+1}) \tau},
\end{aligned}
\end{equation}

\begin{equation} \label{W_forecsts}
\begin{aligned}
\widehat {\mathbf{W}}_{t+1} &= \widehat {W} \left( \tau, u_j, \widehat {\operatorname{UFR}}_{t+1} \right)\\
&= e^{-\ln(1+ \widehat {\operatorname{UFR}}_{t+1}) \left(\tau+u_j\right)} H\left(\tau, u_j, \alpha_{t}\right),
\end{aligned}
\end{equation}
where $\alpha_t$ refers to the value of convergence parameter $\alpha$ at time $t$. For the ZJW method, $\alpha$ is dynamic and changes over time. In contrast, for the SFR and SYC methods, $\alpha$ is set to 0.1 according to the specifications in \cite{Kort2016}. And the $H\left(\tau, u_j, \alpha_{t}\right)$ is defined as follows,
\begin{equation}
H\left(\tau, u_j, \alpha_{t}\right)=\left\{\alpha_{t} \cdot \min \left(\tau, u_{j}\right)-0.5 e^{-\alpha_{t} 	\max \left(\tau, u_{j}\right)}\left(e^{\alpha_{t} \cdot \min 	\left(\tau, u_{j}\right)}-e^{-\alpha_{t} \cdot \min \left(\tau, 	u_{j}\right)}\right)\right\}.
\end{equation}

Finally, the predicted bond price $\widehat {\mathbf{m}}_{t+1}$ is,
\begin{equation} \label{predict_p}
\widehat {\mathbf{m}}_{t+1}=\widehat {\mathbf{q}}_{t+1}+\boldsymbol{\xi}_t \widehat {\mathbf{W}}_{t+1}, 
\end{equation}
and the predicted bond yields can be obtained,
\begin{equation} \label{predict_y}
\begin{aligned}
\widehat {\mathbf{y}}_{t+1} = - \frac{1}{{\bm{\tau}}} \ln \left(\widehat {\mathbf{m}}_{t+1} \right).
\end{aligned}
\end{equation}

In summary, based on \hyperref[ufr_forecasts]{Eq. (\ref{ufr_forecasts})}, \hyperref[q_forecsts]{Eq. (\ref{q_forecsts})}, \hyperref[W_forecsts]{Eq. (\ref{W_forecsts})}, \hyperref[predict_p]{Eq. (\ref{predict_p})}, \hyperref[predict_y]{Eq. (\ref{predict_y})}, we have developed a UFR-based bond yield forecasting model.

We selected the models with positive $R_{\text{oos}}^2$ values from \hyperref[table:oosr2_macro_yields]{Table \ref{table:oosr2_macro_yields}}. For the neural network models, we chose the best-performing model within each category of neural networks, based on its performance in forecasting $\operatorname{\Delta UFR}$, as a representative. Using these models, we constructed the $\operatorname{UFR_{SFR}}$-based bond yield forecasting model, the $\operatorname{UFR_{SYC}}$-based bond yield forecasting model, and the $\operatorname{UFR_{ZJW}}$-based bond yield forecasting model. The $R_{\text{oos}}^2$ values for the predictions of bond yields at different maturities by these three models, compared with the benchmark random walk model, are presented in \hyperref[table:oosr2_ufrbased_yields]{Table \ref{table:oosr2_ufrbased_yields}}.

From \hyperref[table:oosr2_ufrbased_yields]{Table \ref{table:oosr2_ufrbased_yields}}, we observe that, except for XGBoost, which failed to achieve a positive $R_{\text{oos}}^2$ in the prediction of one-year treasury bond yields, all other models demonstrated positive $R_{\text{oos}}^2$ values across all bond maturities, indicating that the bond yield forecasting models based on the predicted $\widehat {\operatorname{UFR}}_{t+1}$ performed well in forecasting bond yields at different maturities. \hyperref[table:oosr2_ufrbased_yields]{Table \ref{table:oosr2_ufrbased_yields}}, when examined horizontally, shows that the predictive performance based on different machine learning models is generally consistent with the results presented in \hyperref[table:oosr2_macro_yields]{Table \ref{table:oosr2_macro_yields}}. Among the linear machine learning models, Ridge and PCR (with 5 components) exhibit the best performance, while neural network models demonstrate the highest performance among the nonlinear models. Vertically, all machine learning models perform exceptionally well in forecasting long-term bond yields, particularly for bonds with maturities exceeding 20 years, where the $R_{\text{oos}}^2$ values are significantly high.

\begin{table}[H]	\caption{\\$R_{\text{oos}}^2$: UFR-Based Bond Yield Prediction vs. Random Walk}
	\setlength{\tabcolsep}{3pt}
	\renewcommand{\arraystretch}{0.95}
	\resizebox{\textwidth}{!}{
		\begin{tabular}{@{}llllllllllll@{}}
			\toprule
			\textbf{Maturity (Yrs)} & 1 & 2 & 3 & 5 & 7 & 10 & 15 & 20 & 30 & 40 & 50 \\ \midrule
			\multicolumn{12}{l}{\textbf{Panel A: $\mathbf{UFR_{SFR}}$-Based Bond Yield Forecasting Model}} \\ \hline
			\textbf{Lasso} & 2.16\%*** & 8.54\%*** & 7.22\%*** & 4.42\%*** & 4.02\%*** & 4.2\%*** & 5.2\%*** & 6.6\%*** & 7.83\%*** & 9.08\%*** & 10.38\%*** \\
			\textbf{Ridge} & 4.73\%*** & 11.01\%*** & 9.71\%*** & 4.83\%*** & 4.28\%*** & 5.69\%*** & 6.92\%*** & 10.31\%*** & 12.03\%*** & 11.57\%*** & 12.74\%*** \\
			\textbf{Elastic net} & 1.7\%*** & 9.05\%*** & 7.85\%*** & 3.6\%*** & 3.07\%*** & 2.91\%*** & 2.82\%*** & 4.15\%*** & 6.92\%*** & 7.98\%*** & 9.23\%*** \\
			\textbf{PCR (5 components)} & 5.06\%*** & 12.81\%*** & 9.88\%*** & 5.73\%*** & 5.83\%*** & 6.66\%*** & 8.82\%*** & 11.23\%*** & 12.09\%*** & 13.38\%*** & 15.58\%*** \\
			\textbf{Random forest} & 2.15\%*** & 8.27\%*** & 6.99\%*** & 3.86\%*** & 4.25\%*** & 5.07\%*** & 7.41\%*** & 8.8\%*** & 9.54\%*** & 11.36\%*** & 12.45\%*** \\
			\textbf{Gradient-boosted trees} & 2.16\%*** & 6.85\%*** & 5.77\%*** & 6.04\%*** & 6.6\%*** & 6.91\%*** & 9.3\%*** & 9.6\%*** & 8.24\%*** & 10.37\%*** & 11.19\%*** \\
			\textbf{XGBoost} & 1.76\%*** & 5.01\%*** & 4.87\%*** & 4.5\%*** & 4.85\%*** & 5.03\%*** & 8.47\%*** & 9.82\%*** & 8.63\%*** & 10.82\%*** & 11.58\%*** \\
			\textbf{Yield-Macro-Net 2 layers} & 4.13\%*** & 7.58\%*** & 8.37\%*** & 7.01\%*** & 8.03\%*** & 12.37\%*** & 14.62\%*** & 14.84\%*** & 15.21\%*** & 15.89\%*** & 17.14\%*** \\
			\textbf{Hybrid net 2   layers} & 3.6\%*** & 10.35\%*** & 9.24\%*** & 4.79\%*** & 4.01\%*** & 6.5\%*** & 6.56\%*** & 9.32\%*** & 10.84\%*** & 10.58\%*** & 13.1\%*** \\
			\textbf{Double net 1   layer} & 2.48\%*** & 4.64\%*** & 3.55\%*** & 2.66\%*** & 4.78\%*** & 6.49\%*** & 8.96\%*** & 11.1\%*** & 11.8\%*** & 14.27\%*** & 16.09\%*** \\
			\textbf{GN-Net  1 layer} & 5.66\%*** & 10.71\%*** & 12\%*** & 7.79\%*** & 8.73\%*** & 7.91\%*** & 10.93\%*** & 11.55\%*** & 15.27\%*** & 15.79\%*** & 17.49\%*** \\ \hline
			\multicolumn{12}{l}{\textbf{Panel B: $\mathbf{UFR_{SYC}}$-Based Bond Yield Forecasting Model}} \\ \hline
			\textbf{Lasso} & 2.39\%*** & 8.91\%*** & 7.51\%*** & 4.69\%*** & 4.39\%*** & 4.48\%*** & 5.59\%*** & 7.01\%*** & 8.14\%*** & 9.56\%*** & 10.84\%*** \\
			\textbf{Ridge} & 4.77\%*** & 10.72\%*** & 9.32\%*** & 4.62\%*** & 4.03\%*** & 5.5\%*** & 6.76\%*** & 10.19\%*** & 11.85\%*** & 11.41\%*** & 12.34\%*** \\
			\textbf{Elastic net} & 2.56\%*** & 10.09\%*** & 8.92\%*** & 4.72\%*** & 4.31\%*** & 4.27\%*** & 4.8\%*** & 6.2\%*** & 8.58\%*** & 9.86\%*** & 10.92\%*** \\
			\textbf{PCR (5 components)} & 5.33\%*** & 13.2\%*** & 10.34\%*** & 6.25\%*** & 6.44\%*** & 7.28\%*** & 9.5\%*** & 11.95\%*** & 12.72\%*** & 14.17\%*** & 16.24\%*** \\
			\textbf{Random forest} & 1.43\%*** & 6.91\%*** & 5.38\%*** & 2.74\%*** & 2.66\%*** & 3.19\%*** & 5.37\%*** & 7.39\%*** & 8.16\%*** & 9.25\%*** & 10.24\%*** \\
			\textbf{Gradient-boosted trees} & 2.12\%*** & 7.15\%*** & 6.26\%*** & 6.61\%*** & 6.84\%*** & 6.78\%*** & 9.04\%*** & 9.49\%*** & 7.86\%*** & 9.88\%*** & 10.62\%*** \\
			\textbf{XGBoost} & 2.78\%*** & 7.06\%*** & 6.71\%*** & 6.18\%*** & 6.21\%*** & 6.43\%*** & 9.5\%*** & 11.06\%*** & 9.44\%*** & 11.46\%*** & 11.93\%*** \\
			\textbf{Yield-Macro-Net 2 layers} & 3.77\%*** & 7.27\%*** & 8.12\%*** & 6.92\%*** & 7.85\%*** & 12.21\%*** & 14.47\%*** & 14.73\%*** & 14.99\%*** & 15.48\%*** & 16.57\%*** \\
			\textbf{Hybrid net 2   layers} & 4.21\%*** & 10.89\%*** & 9.95\%*** & 5.61\%*** & 5.02\%*** & 7.36\%*** & 7.5\%*** & 10.32\%*** & 11.76\%*** & 11.55\%*** & 13.97\%*** \\
			\textbf{Double net 1   layer} & 2.47\%*** & 4.77\%*** & 3.53\%*** & 2.6\%*** & 4.58\%*** & 6.29\%*** & 8.67\%*** & 10.87\%*** & 11.64\%*** & 14.02\%*** & 15.88\%*** \\
			\textbf{GN-Net  1 layer} & 6.15\%*** & 10.97\%*** & 12.14\%*** & 7.75\%*** & 8.66\%*** & 7.64\%*** & 10.55\%*** & 11.37\%*** & 15.29\%*** & 15.81\%*** & 17.46\%*** \\ \hline
			\multicolumn{12}{l}{\textbf{Panel C: $\mathbf{UFR_{ZJW}}$-Based Bond Yield Forecasting Model}} \\ \hline
			\textbf{Lasso} & 2.98\%*** & 8.36\%*** & 7.39\%*** & 5.45\%*** & 5.65\%*** & 6.04\%*** & 6.51\%*** & 7.11\%*** & 8.51\%*** & 10.23\%*** & 10.96\%*** \\
			\textbf{Ridge} & 5.37\%*** & 11.13\%*** & 10.12\%*** & 7.33\%*** & 7.07\%*** & 8.73\%*** & 9.23\%*** & 12.07\%*** & 13.4\%*** & 13.72\%*** & 14.09\%*** \\
			\textbf{Elastic net} & 2.23\%*** & 8\%*** & 7.71\%*** & 5.29\%*** & 5.23\%*** & 5.61\%*** & 4.87\%*** & 6.01\%*** & 9.01\%*** & 10.51\%*** & 11.46\%*** \\
			\textbf{PCR (5 components)} & 5.5\%*** & 12.55\%*** & 10.22\%*** & 7.47\%*** & 7.98\%*** & 8.78\%*** & 10.38\%*** & 12.13\%*** & 12.91\%*** & 15\%*** & 16.8\%*** \\
			\textbf{Random forest} & 1.91\%*** & 6.01\%*** & 5.5\%*** & 3.61\%*** & 3.98\%*** & 4.66\%*** & 7.12\%*** & 8.81\%*** & 9.09\%*** & 10.48\%*** & 11.08\%*** \\
			\textbf{Gradient-boosted trees} & 1.21\%*** & 6.74\%*** & 4.69\%*** & 4.24\%*** & 4.7\%*** & 4.42\%*** & 5.16\%*** & 6.97\%*** & 8.96\%*** & 9.84\%*** & 10.67\%*** \\
			\textbf{XGBoost} & -0.21\% & 3.99\%*** & 1.84\%*** & 1.61\%*** & 0.89\%*** & 1.48\%*** & 2.26\%*** & 5.27\%*** & 6.31\%*** & 5.42\%*** & 5.98\%*** \\
			\textbf{Yield-Macro-Net 2 layers} & 5.79\%*** & 9.54\%*** & 9.83\%*** & 7.84\%*** & 9.88\%*** & 15.13\%*** & 15.32\%*** & 16.5\%*** & 18.19\%*** & 17.89\%*** & 19.95\%*** \\
			\textbf{Hybrid net 2   layers} & 4.76\%*** & 11.18\%*** & 10.33\%*** & 7.29\%*** & 6.91\%*** & 9.46\%*** & 9.73\%*** & 12.36\%*** & 13.24\%*** & 13.26\%*** & 14.68\%*** \\
			\textbf{Double net 1   layer} & 3.16\%*** & 6.12\%*** & 5.38\%*** & 4.69\%*** & 7.26\%*** & 9.26\%*** & 11.54\%*** & 13.35\%*** & 13.77\%*** & 16.03\%*** & 17.01\%*** \\
			\textbf{GN-Net  1 layer} & 8.53\%*** & 14.01\%*** & 15.68\%*** & 13.09\%*** & 14.95\%*** & 14.35\%*** & 16.16\%*** & 17.26\%*** & 22.02\%*** & 22.11\%*** & 22.94\%*** \\
			\bottomrule
	\end{tabular}}
	\begin{tablenotes}
		\fontsize{10}{10}\selectfont
		\item[1] Note: This table presents the out-of-sample $R_{\text{oos}}^2$ of bond yield predictions based on the UFR for the t+1 rolling forward forecast. CW statistics are annotated with ``*'', ``**'', and ``***'' to indicate significance at the 10\%, 5\%, and 1\% levels. Significance levels are reported only when $R_{\text{oos}}^2 > 0$. 
	\end{tablenotes}
	\label{table:oosr2_ufrbased_yields}
\end{table}

For the neural network models, the $R_{\text{oos}}^2$ for forecasting bond yields with maturities greater than 30 years exceeds 10\%. In contrast, the prediction of one-year treasury bond yields performs relatively weakly, exhibiting the lowest performance across all maturities. It is noteworthy that the performance in forecasting bond yields at different maturities does not increase monotonically with maturity length; rather, it follows a wave-like pattern, initially increasing, then decreasing, before rising again.

\section{Conclusions}
This study integrates bond yields and macroeconomic variables, employing both linear and nonlinear machine learning methods to predict the UFR and enhance the interpretability of the model through the incorporation of macroeconomic factors. By utilizing data from Chinese treasury bonds and macroeconomic variables between December 2009 and December 2024, we demonstrate the effectiveness of de Kort-Vellekoop-type methods in estimating the UFR, highlighting their capacity to capture underlying trends and dynamics.

Furthermore, we apply these methods to identify the optimal turning parameter within the ZJW improved method, which plays a crucial role in smoothing out the UFR’s anomalous fluctuations. This adjustment significantly improves the stability and reliability of the UFR estimation, providing more consistent and accurate forecasts.

Our findings further emphasize the superior performance of nonlinear machine learning techniques in predicting the UFR and ultra-long-term bond yields compared to linear models. Nonlinear models capture complex relationships between variables more effectively, leading to better predictive outcomes. Moreover, the incorporation of macroeconomic variables, especially those related to price indices, markedly enhances the forecasting accuracy of nonlinear models. This underscores the importance of considering macroeconomic factors, as they provide valuable context and improve the robustness of the predictive models.

Lastly, based on the predicted UFR, we propose novel UFR-based bond yield forecasting models. The empirical results demonstrate the exceptional performance of these models in predicting bond yields across different maturities. This UFR-based bond yield forecasting model fills a gap in the application of UFR for predicting the term structure of bond yields, making a significant contribution to the application direction of UFR forecasting.

In conclusion, the results of this study contribute to a deeper understanding of the UFR prediction process and underscore the significance of integrating both machine learning techniques and macroeconomic variables in financial forecasting. The findings also provide valuable insights for policymakers and financial analysts seeking more reliable tools for forecasting long-term interest rates and managing economic uncertainties. More importantly, the proposed UFR-based bond yield term structure forecasting model presents bond investors with a promising new approach to predicting the term structure of bond yields and paves the way for expanded applications of the UFR.

%
%
%
%

\bibliographystyle{elsarticle-harv}
\bibliography{references}

\begin{appendices}
	
\renewcommand\thefigure{\Alph{section}\arabic{figure}}   
\renewcommand\thetable{\Alph{section}\arabic{table}} 
\renewcommand\theequation{\Alph{section}\arabic{equation}}
\setcounter{equation}{0}
\setcounter{figure}{0}  
\setcounter{table}{0}	

\section{List of Abbreviations}


\begin{table}[H]
	\caption{\\List of abbreviations}
	\setlength{\tabcolsep}{5pt}
	\renewcommand{\arraystretch}{1}
	\resizebox{\textwidth}{!}{
	\begin{tabular}{@{}lll@{}}
		\toprule
		\textbf{No.} & \textbf{Abbreviation}     & \textbf{Full Name}             \\ \midrule
		\textbf{1}   & \textbf{UFR}     & Ultimate Forward Rate          \\
		\textbf{2}  & \textbf{EIOPA}     & European Insurance and Occupational Pensions Authority                                \\
		\textbf{3}   & \textbf{DNS}     & Dynamic Nelson-Siegel          \\
		\textbf{4}   & \textbf{SDF}     & Smoothest Discount Factor      \\
		\textbf{5}   & \textbf{NS}      & Nelson-Siegel                  \\
		\textbf{6}   & \textbf{ZJW}     & Zhao, Jia and Wu               \\
		\textbf{7}   & \textbf{SFR}     & Smoothest Forward Rate         \\
		\textbf{8}   & \textbf{SYC}     & Smoothest Yield Curve          \\
		\textbf{9}   & \textbf{OLS}     & Ordinary Least Squares         \\
		\textbf{10}  & \textbf{PCR}     & Principal Component Regression \\
		\textbf{11}  & \textbf{PLS}     & Partial Least Squares          \\
		\textbf{12}  & \textbf{EN}      & Elastic Net                    \\
		\textbf{13}  & \textbf{RT}      & Regression Trees               \\
		\textbf{14} & \textbf{GBRT}      & Gradient-Boosting Regression Trees                                                    \\
		\textbf{15}  & \textbf{XGBoost} & Extreme Gradient Boosting      \\
		\textbf{16}  & \textbf{NN}      & Neural Networks                \\
		\textbf{17}  & \textbf{PC}      & Principal Component            \\
		\textbf{18}  & \textbf{MLP}     & Multilayer Perceptrons         \\
		\textbf{19}  & \textbf{GN-Net}  & Group Ensemble Net             \\
		\textbf{20}  & \textbf{RMSE}    & Root Mean Squared Error        \\
		\textbf{21}  & \textbf{MAE}     & Mean Absolute Error            \\
		\textbf{22}  & \textbf{CW}      & Clark and West                 \\
		\textbf{23}  & \textbf{PCA}     & Principal Component Analysis   \\
		\textbf{24}  & \textbf{SHAP}    & SHapley Additive exPlanations  \\
		\textbf{25} & \textbf{RDRRSMDFI} & RMB Deposit Reserve Ratio for   Small and Medium-Sized Deposit Financial Institutions \\ \bottomrule
	\end{tabular}}
\end{table}

\renewcommand\thefigure{\Alph{section}\arabic{figure}}   
\renewcommand\thetable{\Alph{section}\arabic{table}} 
\renewcommand\theequation{\Alph{section}\arabic{equation}}
\setcounter{equation}{0}
\setcounter{figure}{0}  
\setcounter{table}{0}	
\section{UFR Determination by SFR and SYC Methods}	\label{ufr_sfr_syc}

In the study conducted by \cite{Kort2016}, the asymptotic forward rates, $f_{\infty}$, are represented as a linear combination of yield values. Specifically:

\begin{equation}
f_{\infty}=\sum_{k=0}^n v_k y_k .
\end{equation}

The weights, $\left\{v_k, k=1, \ldots, n\right\}$, are defined as $v_k=\sum_{j=1}^n\left[G^{-1}\right]_{j k}$, with the additional condition that $v_0=1-\sum_{k=1}^n v_k$. Here, the matrix $\mathbf{G}$ represents either the SFR or SYC method and is given by:

\begin{equation}
G_{k j}^f=\frac{1}{u_k} \int_0^{u_k} \bar{W}\left(s, u_j\right) d s, \quad G_{k j}^y=\frac{1}{\alpha u_j} W\left(u_k, u_j\right) .
\end{equation}

In these expressions, $W\left(u_k, u_j\right)$ is defined in \hyperref[W_and_H]{Eq. (\ref{W_and_H})}, and $\bar{W}(\tau, u)$ is formulated as:

\begin{equation}
\bar{W}(\tau, u)=1-e^{-\bar{\alpha} \tau} \frac{\cosh (\bar{\alpha} u)-1}{\frac{1}{2} \bar{\alpha}^2 u^2}+\mathbf{1}_{\tau \leq u}\left(\frac{\cosh (\bar{\alpha}(u-\tau))-1-\frac{1}{2} \bar{\alpha}^2(u-\tau)^2}{\frac{1}{2} \bar{\alpha}^2 u^2}\right).
\end{equation}

These functions are derived as affine combinations of integrals of Smith-Wilson functions, resulting in a smoother form compared to the original Smith-Wilson functions.

If the cashflow matrix $\mathbf{C}_{i j}$ is invertible, the yields can be directly observed since $\pi=\mathbf{C}^{-1} \mathbf{m}$ and $y_k=-\frac{\ln \pi_k}{u_k}$. In this case, both $f_{\infty}^f$ and $f_{\infty}^y$ can be computed as linear combinations of quantities observable in the market, with fixed weights that are predetermined for a specific set of maturities $\left\{u_i\right\}_{i \in \ell}$.

When the short rate is not specified in advance, it can be treated as a free parameter in the optimization process. An explicit formula for this short rate, which ensures the smoothest curve for small maturities, can be derived when the cashflow matrix is invertible.

In cases where $n_{\ell }= n_{\mathscr{G}} = n,\left(c_{i j}\right)_{i \in \ell, j \in \mathscr{G}}$ and the matrix $\left(c_{i j}\right)_{i \in \ell, j \in \mathscr{G}}$ is invertible, and the initial short rate is unknown, the previously mentioned formula remains valid. The unknown short rate can be substituted by its optimized value, as follows:

\begin{equation}
y_0=\frac{\sum_{j=1}^n \frac{1}{u_j} \sum_{k=1}^n G_{j k}^{-1} \frac{y\left(u_j\right)+y\left(u_k\right)}{2}}{\sum_{j=1}^n \frac{1}{u_j} \sum_{k=1}^n G_{j k}^{-1}} .
\end{equation}

In this case, $\mathbf{G}$ represents either $\mathbf{G}^f$ or $\mathbf{G}^y$, as defined earlier.

\renewcommand\thefigure{\Alph{section}\arabic{figure}}   
\renewcommand\thetable{\Alph{section}\arabic{table}} 
\renewcommand\theequation{\Alph{section}\arabic{equation}}
\setcounter{equation}{0}
\setcounter{figure}{0}  
\setcounter{table}{0}

\section{Macroeconomic Variables}	\label{ufr:abbr_macro}

\setlength{\tabcolsep}{0pt}
\small
\renewcommand{\arraystretch}{0.65}
\begin{longtable}{@{\extracolsep{\fill}}>{\bfseries}c>{\bfseries}c>{\raggedright\arraybackslash}p{0.8\textwidth}@{}}

	\caption{List of Macroeconomic Variables} \label{tab:macro_vars} \\
	\toprule
	\textbf{No.} & \textbf{Abbr.} & \textbf{Name of macroeconomic variable} \\
	\midrule
	\endfirsthead
	\multicolumn{3}{c}{{\bfseries Table \thetable\ Continued}} \\
	\toprule
	\textbf{No.} & \textbf{Abbr.} & \textbf{Name of macroeconomic variable} \\
	\midrule
	\endhead
	\midrule
	\multicolumn{3}{r}{{Continued on next page}} \\
	\endfoot
	\bottomrule
	\endlastfoot
\multicolumn{3}{l}{\textbf{Panel A.   Macro-prosperity index}}                                              \\
\textbf{1}   & \textbf{MICI}      & Macroeconomic Index: Coincident Index                                   \\
\textbf{2}   & \textbf{MILI\_1}   & Macroeconomic Index: Leading Index                                      \\
\textbf{3}   & \textbf{MILI\_2}   & Macroeconomic Index: Lagging Index                                      \\ \midrule
\multicolumn{3}{l}{\textbf{Panel B. Output}}                                                                \\ \midrule
\textbf{4}   & \textbf{IAVAD}    & Industrial Added Value: Above   Designated Size Industrial Enterprises: YoY \\
\textbf{5}   & \textbf{MPMI}      & Manufacturing PMI                                                       \\ \midrule
\multicolumn{3}{l}{\textbf{Panel C.   Consumption \& Retail}}                                               \\ \midrule
\textbf{6}   & \textbf{CCI}       & Consumer Confidence Index                                               \\
\textbf{7}   & \textbf{CCIS}      & Consumer Confidence Index: Satisfaction                                 \\
\textbf{8}   & \textbf{CCIE}      & Consumer Confidence Index: Expectations                                 \\
\textbf{9}   & \textbf{TRS}       & Total Retail Sales of Social Consumer Goods: YoY                        \\
\textbf{10}  & \textbf{TRSRG}     & Total Retail Sales of Social Consumer Goods: Retail Goods: YoY          \\
\textbf{11}  & \textbf{TRSCR}     & Total Retail Sales of Social Consumer Goods: Catering Revenue:   YoY    \\
\textbf{12}  & \textbf{RSGOFBTA}  & Retail Sales: Grains, Oil, Food, Beverages, Tobacco \&   Alcohol: YoY   \\
\textbf{13}  & \textbf{RSGOF}     & Retail Sales: Grains, Oil, Food: YoY                                    \\
\textbf{14}  & \textbf{RSSB}      & Retail Sales: Beverages: YoY                                            \\
\textbf{15}  & \textbf{RSTA}      & Retail Sales: Tobacco \& Alcohol: YoY                                   \\
\textbf{16}  & \textbf{RSAF}      & Retail Sales: Apparel, Footwear, Textiles: YoY                          \\
\textbf{17}  & \textbf{RSA}       & Retail Sales: Apparel: YoY                                              \\
\textbf{18}  & \textbf{RSC}       & Retail Sales: Cosmetics: YoY                                            \\
\textbf{19}  & \textbf{RSGJ}      & Retail Sales: Gold, Silver, Jewelry: YoY                                \\
\textbf{20}  & \textbf{RSG}       & Retail Sales: Daily Goods: YoY                                          \\
\textbf{21}  & \textbf{RSUASEG}   & Retail Sales: Sports, Entertainment Goods: YoY                          \\
\textbf{22}  & \textbf{RSBN}      & Retail Sales: Books, Newspapers, Magazines: YoY                         \\
\textbf{23}  & \textbf{RSHA}      & Retail Sales: Household Appliances and Audio-Visual Equipment:   YoY    \\
\textbf{24}  & \textbf{RSCWM}     & Retail Sales: Chinese and Western Medicines: YoY                        \\
\textbf{25}  & \textbf{RSCO}      & Retail Sales: Cultural and Office Supplies: YoY                         \\
\textbf{26}  & \textbf{RSF}       & Retail Sales: Furniture: YoY                                            \\
\textbf{27}  & \textbf{RSCE}      & Retail Sales: Communications Equipment: YoY                             \\
\textbf{28}  & \textbf{RSPP}      & Retail Sales: Petroleum and Products: YoY                               \\
\textbf{29}  & \textbf{RSBD}      & Retail Sales: Building and Decoration Materials: YoY                    \\
\textbf{30}  & \textbf{RSA}       & Retail Sales: Automobiles: YoY                                          \\ \midrule
\multicolumn{3}{l}{\textbf{Panel D. Price   index}}                                                         \\ \midrule
\textbf{31}  & \textbf{CPI}       & CPI: YoY                                                                \\
\textbf{32}  & \textbf{CPIF}      & CPI: Food: YoY                                                          \\
\textbf{33}  & \textbf{CPINF}     & CPI: Non-Food: YoY                                                      \\
\textbf{34}  & \textbf{CPICG}     & CPI: Consumer Goods: YoY                                                \\
\textbf{35}  & \textbf{CPIS}      & CPI: Services: YoY                                                      \\
\textbf{36}  & \textbf{CPIC}      & CPI: Clothing: YoY                                                      \\
\textbf{37}  & \textbf{CPIH}      & CPI: Housing: YoY                                                       \\
\textbf{38}  & \textbf{CPIHGS}    & CPI: Household Goods and Services: YoY                                  \\
\textbf{39}  & \textbf{CPITC}     & CPI: Transportation and Communication: YoY                              \\
\textbf{40}  & \textbf{CPIECE}    & CPI: Education, Culture, and Entertainment: YoY                         \\
\textbf{41}  & \textbf{CPIHC}     & CPI: Health Care: YoY                                                   \\
\textbf{42}  & \textbf{PPIAIP}    & PPI: All Industrial Products: YoY                                       \\
\textbf{43}  & \textbf{PPIPG}     & PPI: Producer Goods: YoY                                                \\
\textbf{44}  & \textbf{PPICT}     & PPI: Consumer Goods: YoY                                                \\
\textbf{45}  & \textbf{EPIHS2}    & Export Price Index (HS2): YoY                                           \\
\textbf{46}  & \textbf{IPIH2}     & Import Price Index (HS2): YoY                                           \\ \midrule
\multicolumn{3}{l}{\textbf{Panel E. Interest rate}}                                                       \\ \midrule
\textbf{47}  & \textbf{DDR}       & Demand Deposit Rate                                                     \\
\textbf{48}  & \textbf{TDR3}      & Time Deposit Rate: 3 Months                                             \\
\textbf{49}  & \textbf{TDR6}      & Time Deposit Rate: 6 Months                                             \\
\textbf{50}  & \textbf{TDRF1}     & Time Deposit Rate (Fixed): 1 Year                                       \\
\textbf{51}  & \textbf{TDRF2}     & Time Deposit Rate (Fixed): 2 Years                                      \\
\textbf{52}  & \textbf{TDRF3}     & Time Deposit Rate (Fixed): 3 Years                                      \\
\textbf{53}  & \textbf{STLR6}     & Short-Term Loan Rate: 6 Months (Inclusive)                              \\
\textbf{54}  & \textbf{STLR6\_1}  & Short-Term Loan Rate: 6 Months to 1 Year (Inclusive)                    \\
\textbf{55}  & \textbf{MLTLR1\_3} & Medium to Long-Term Loan Rate: 1 to 3 Years (Inclusive)                 \\
\textbf{56}  & \textbf{MLTLR3\_5} & Medium to Long-Term Loan Rate: 3 to 5 Years (Inclusive)                 \\
\textbf{57}  & \textbf{MLTLR5\_}  & Medium to Long-Term Loan Rate: Over 5 Years                             \\ \midrule
\multicolumn{3}{l}{\textbf{Panel F. Money   \& Credit}}                                                     \\ \midrule
\textbf{58}  & \textbf{M0}        & M0: YoY                                                                 \\
\textbf{59}  & \textbf{M1}        & M1: YoY                                                                 \\
\textbf{60}  & \textbf{M2}        & M2: YoY                                                                 \\
\textbf{61}  & \textbf{ORAFER}    & Official Reserve Assets: Foreign Exchange Reserves                      \\
\textbf{62}  & \textbf{ORAGP}     & Official Reserve Assets: Gold (Measured in Pure Gold Ounces)            \\
\textbf{63}  & \textbf{FID}       & Financial Institutions Deposit Balance: RMB                             \\
\textbf{64}  & \textbf{FIDY}      & Financial Institutions Deposit Balance: RMB: YoY                        \\
\textbf{65}  & \textbf{FIDRE}     & Financial Institutions Deposit Balance: Enterprises: RMB                \\
\textbf{66}  & \textbf{FIDRGA}    & Financial Institutions Deposit Balance: Government Agencies:   RMB      \\
\textbf{67}  & \textbf{FIDRF}     & Financial Institutions Deposit Balance: Fiscal: RMB                     \\
\textbf{68}  & \textbf{FIDRS}     & Financial Institutions Deposit Balance: Savings: RMB                    \\
\textbf{69}  & \textbf{FILRMB}    & Financial Institutions Loan Balance: RMB                                \\
\textbf{70}  & \textbf{FILRMBY}   & Financial Institutions Loan Balance: RMB: YoY                           \\
\textbf{71}  & \textbf{CIHUSTB}   & Chinese Investors Holding US Treasury Bonds                             \\ \midrule
\multicolumn{3}{l}{\textbf{Panel G. Investment}}                                                            \\ \midrule
\textbf{72}  & \textbf{FAIC}      & Fixed Asset Investment Completion:   Cumulative YoY                     \\
\textbf{73}  & \textbf{FAICPI}    & Fixed Asset Investment Completion: Primary Industry:   Cumulative YoY   \\
\textbf{74}  & \textbf{FAICSI}    & Fixed Asset Investment Completion: Secondary Industry:   Cumulative YoY \\
\textbf{75}  & \textbf{FAICTI}    & Fixed Asset Investment Completion: Tertiary Industry:   Cumulative YoY  \\ \midrule
\multicolumn{3}{l}{\textbf{Panel H. Real   estate}}                                                         \\ \midrule
\textbf{76}  & \textbf{REDIC}     & Real Estate Development Investment   Completion: Cumulative YoY         \\
\textbf{77}  & \textbf{CHSAC}     & Commercial Housing Sales Area: Cumulative YoY                           \\
\textbf{78}  & \textbf{NSHAC}     & Newly Started Housing Area: Cumulative YoY                              \\ \midrule
\multicolumn{3}{l}{\textbf{Panel I. Tax}}                                                                   \\ \midrule
\textbf{79}  & \textbf{TRC}       & Tax Revenue: Cumulative YoY                                             \\
\textbf{80}  & \textbf{GPBR}      & General Public Budget Revenue: Cumulative YoY                           \\
\textbf{81}  & \textbf{GPBE}      & General Public Budget Expenditure: Cumulative YoY                       \\ \midrule
\multicolumn{3}{l}{\textbf{Panel J. Trade}}                                                                 \\ \midrule
\textbf{82}  & \textbf{IEA}       & Import and Export Amount: YoY                                           \\
\textbf{83}  & \textbf{EA}        & Export Amount: YoY                                                      \\
\textbf{84}  & \textbf{IA}        & Import Amount: YoY                                                      \\
\textbf{85}  & \textbf{TB}        & Trade Balance: YoY                                                      \\ \midrule
\multicolumn{3}{l}{\textbf{Panel K. Foreign   exchange rate}}                                               \\ \midrule
\textbf{86}  & \textbf{REERI}     & Real Effective Exchange Rate Index:   RMB: Broad Measure                \\
\textbf{87}  & \textbf{NEERI}     & Nominal Effective Exchange Rate Index: RMB: Broad Measure               \\
\textbf{88}  & \textbf{AERUSD}    & Average Exchange Rate: USD to RMB                                       \\
\textbf{89}  & \textbf{AEREUR}    & Average Exchange Rate: EUR to RMB                                       \\
\textbf{90}  & \textbf{AERHKD}    & Average Exchange Rate: HKD to RMB                                       \\
\textbf{91}  & \textbf{AERJPY}    & Average Exchange Rate: 100 JPY to RMB                                   \\ \midrule
\multicolumn{3}{l}{\textbf{Panel L. Stock   market}}                                                        \\ \midrule
\textbf{92}  & \textbf{SSE50}     & SSE 50 Index                                                            \\
\textbf{93}  & \textbf{CSI300}    & CSI 300 Index                                                           \\
\textbf{94}  & \textbf{CSI500}    & CSI 500 Index                                                           \\
\textbf{95}  & \textbf{SZSESCI}   & Shenzhen Stock Exchange: Shenzhen Composite Index                       \\
\textbf{96}  & \textbf{SSECI}     & Shanghai Stock Exchange Composite Index                                 \\
\textbf{97}  & \textbf{SSABTV}    & Shanghai/Shenzhen Stock Markets: A/B Shares: Total Market   Value       \\
\textbf{98}  & \textbf{SSABCV}    & Shanghai/Shenzhen Stock Markets: A/B Shares: Circulating   Market Value \\
\textbf{99}  & \textbf{SSSTV}     & Shanghai/Shenzhen Stock Markets: Stock Trading Volume                   \\
\textbf{100} & \textbf{SSSTQ}     & Shanghai/Shenzhen Stock Markets: Stock Trading Quantity                 \\
\textbf{101} & \textbf{SSDFFA}   & Shanghai/Shenzhen Stock Markets: Domestic and Foreign   Fundraising Amount  \\ \midrule
\multicolumn{3}{l}{\textbf{Panel M. Monetary   policy}}                                                     \\ \midrule
\textbf{102} & \textbf{DRRLI}     & RMB Deposit Reserve Ratio: Large   Institutions                         \\
\textbf{103} & \textbf{DRRSMFI}  & RMB Deposit Reserve Ratio: Small and Medium-Sized Deposit   Institutions    \\
\textbf{104} & \textbf{PBoCLRDR} & People’s Bank of China Loan Rate to Financial Institutions:   Discount Rate \\
\textbf{105} & \textbf{SHIBORO}   & SHIBOR: Overnight                              \\ 
\end{longtable}
\normalsize 
\renewcommand{\arraystretch}{1}

\end{appendices}	

\end{document}